\documentclass[journal,twocolumn]{IEEEtran}
\usepackage{ifpdf}
\usepackage[most]{tcolorbox}
\usepackage{cite}
\usepackage{graphicx}
\ifCLASSINFOpdf
  \DeclareGraphicsExtensions{.pdf,.png,.jpg,.jpeg}
\else
  \DeclareGraphicsExtensions{.eps}
\fi
\usepackage{subcaption}  % Add this in the preamble

\ifCLASSOPTIONcompsoc
\usepackage[caption=false, font=scriptsize, labelfont=sf, textfont=sf]{subfig}
\else
\usepackage[caption=false, font=footnotesize]{subfig}
\fi
\usepackage{floatrow}
\usepackage{algorithm,algorithmic} % for algorithm
\usepackage{caption}
\usepackage{graphicx} % Required for inserting images
\usepackage{amsmath,amsfonts,bm,amssymb,bbm,dsfont,amsthm}
\usepackage{mathalfa}
\usepackage[thinc]{esdiff}
\usepackage[normalem]{ulem}
\usepackage{ragged2e}
\usepackage{caption}
\usepackage{color}
\usepackage{subfig}
\newcommand{\bs}[1]{\boldsymbol{#1}}
\newcommand{\mc}[1]{\mathcal{#1}}

\newcommand{\mb}[1]{\mathbf{#1}}
\newcommand{\mr}[1]{\mathrm{#1}}

\newcommand{\e}{\mathrm{e}}
\newcommand{\ji}{\mathrm{j}}

\newcommand{\s}{\mathrm{s}}
\newcommand{\E}{\mathbb{E}}
\newcommand{\lr}[1]{\langle #1 \rangle}
\newcommand{\blr}[1]{\big\langle #1 \big\rangle}
\newcommand{\bblr}[1]{\bigg\langle #1 \bigg\rangle}

\newcommand{\C}{\mathbb{C}}

\newcommand{\RR}{\mathbb{R}}

\newtheorem{Remark}{\textbf{Remark}}

\usepackage[normalem]{ulem}

\begin{document}
	
	\title{Semi-Gridless Variational Bayes Channel Estimation in XL-MIMO: Near-Field Modeling and Inference}

        \author{
    \IEEEauthorblockN{
        Van-Chung Luu\IEEEauthorrefmark{1}\IEEEauthorrefmark{2}, 
        Toan-Van Nguyen\IEEEauthorrefmark{1}, 
        Nuria González-Prelcic\IEEEauthorrefmark{2}, 
        and Duy H. N. Nguyen\IEEEauthorrefmark{1}
    }\\
    \IEEEauthorblockA{
    \begin{tabular}{l}
        \IEEEauthorrefmark{1}Department of Electrical and Computer Engineering, San Diego State University, San Diego, CA, USA\\
        \IEEEauthorrefmark{2}Department of Electrical and Computer Engineering, University of California, San Diego, CA, USA
    \end{tabular}
}

        \thanks{Part of this work has been published at the 2025 Asilomar Conference on Signals, Systems, and Computers, Pacific Grove, CA, USA, Oct. 2025 \cite{conf}.}
	
	% \IEEEauthorblockA{Emails: 
	% 	%\IEEEauthorrefmark{1}
 %        \{cluu1171, tnguyen58,  duy.nguyen\}@sdsu.edu} %\IEEEauthorrefmark{1}duy.nguyen@sdsu.edu}
  \vspace{-1.8em}
}

\maketitle

\begin{abstract}
Extremely large antenna arrays and high-frequency operation are two key technologies that advance performance metrics such as higher data rates, lower latency, and wider coverage in sixth-generation  communications. However, the adoption of these technologies fundamentally changes the characteristics of wavefronts, forcing communication systems to operate in the near-field region. The transition from planar far-field communications to spherical near-field propagation necessitates novel channel estimation algorithms to fully exploit the unique features of spherical wavefronts for advanced transceiver design. To this end, we propose a novel semi-gridless channel estimation approach based on a variational Bayesian (VB) inference framework. Specifically, we reformulate the near-field channel model for both uniform linear arrays and uniform planar arrays into separate direction-of-arrival (DoAs) and distance components. Building on these new representations, we employ a gridless approach for DoAs estimation using a von Mises distribution, and a coarse-to-fine grid search for distance estimation. We then develop a semi-gridless variational Bayesian (SG-VB) algorithm with efficient update rules that enables accurate channel reconstruction. Simulation results validate the effectiveness of the proposed SG-VB algorithm, demonstrating enhanced near-field channel reconstruction accuracy and superior estimation performance for both DoAs and distance components embedded in near-field channels.
\end{abstract}

\section{Introduction}
Sixth-generation (6G) communications are expected to support extremely high data rates, potentially on the order of Tb/s, together with low latency and enhanced service capabilities \cite{wang2023road}. These performance requirements have motivated interest in the use of extremely large antenna arrays (ELAAs), which offer increased spatial degrees of freedom and improved spectral efficiency \cite{bjornson2019massive, ye2024extremely}. Furthermore, there is an increasing shift toward utilizing millimeter-wave and terahertz bands to leverage the expansive bandwidths necessary for high-speed transmission\cite{thomas2025survey}. The use of ELAAs and high-frequency operation improves the spectral efficiency and spatial resolution of wireless networks \cite{an2024near}, but necessitates more accurate electromagnetic propagation models that capture wavefront curvature and near-field effects \cite{liu2023near}. Specifically, the electromagnetic radiation field can be roughly divided into two regions—near-field and far-field \cite{liu2023near, lu2024tutorial}—with the boundary defined by the Rayleigh distance \cite{balanis2016antenna}. The Rayleigh distance is proportional to the square of the array aperture and inversely proportional to the wavelength. As the antenna aperture and carrier frequency increase, the Rayleigh distance grows and can reach hundreds of meters, shifting the near–far field boundary such that many communication links operate in the near-field of the base station (BS).

% As antenna aperture and carrier frequency increase, the Rayleigh distance can extend to hundreds of meters, resulting in the operation of communication networks in the near-field region of the base station (BS)

Unlike conventional far-field systems, where planar wavefronts are characterized primarily by the angle of arrival (AoA), near-field extra-large multiple-input multiple-output (XL-MIMO) requires modeling spherical wavefronts whose phase varies with the element-dependent propagation distance from the user to each antenna element\cite{liu2023near, liu2025near}. On the one hand, this opens new opportunities and services for both communication and sensing networks \cite{liu2023near}, such as offering greater multiplexing gains and higher-resolution user location estimation. On the other hand, it necessitates novel channel estimation (CE) algorithms, as conventional far-field CE methods based on the angular domain are no longer suitable. 

% The key distinction between the far-field and near-field lies in the curvature of the propagation waves. In the far-field region, the planar-wave channel model is applicable, resulting in a linear signal phase. In contrast, when networks operate in the near-field region, spherical-wave channel models must be used to accurately capture wireless channel characteristics \cite{liu2023near, liu2025near}. The transition from planar-wave to spherical-wave propagation introduces a new distance dimension

%\textbf{Subspace-based methods}

To achieve high estimation accuracy in near-field channels, traditional subspace-based methods, such as the Multiple Signal Classification (MUSIC) algorithm \cite{stoica2005spectral}, have been extensively investigated. For example, two-dimensional (2D) MUSIC for direction-of-arrival (DoA) estimation in mixed far-field and near-field scenarios has been examined in \cite{he2011efficient}, and its application to channel estimation in near-field XL-MIMO systems \cite{qu2024two, kosasih2023parametric, 10908613} has demonstrated the effectiveness of subspace-based methods. In addition, the authors of \cite{huang2023low} proposed a sequential angle–distance channel estimation method, which achieves performance comparable to MUSIC but with lower complexity. Despite their high-resolution estimation capability, these methods suffer from the high complexity of exhaustive search and the large number of samples required to overcome rank deficiency \cite{kosasih2023parametric}, which hinders their application in ELAAs-MIMO systems.

%\textbf{Codebook-based methods}

% Addressing the limitations of subspace-based methods, researchers have turned to codebook-based techniques
To overcome the limitations of subspace-based methods, recent research has increasingly focused on codebook-based techniques. The main idea is to exploit the sparsity of the channel in the polar domain by constructing a codebook from sampled angles and distances \cite{cui2022channel, wu2023multiple, guo2023compressed, zhang2023codebook}. Based on this codebook, compressive sensing techniques are employed to efficiently recover the channel. In \cite{cui2022channel}, polar-domain simultaneous orthogonal matching pursuit (P-SOMP) and polar-domain simultaneous iterative gridless weighted (P-SIGW) methods were proposed. However, the compressive sensing-based approaches do not explicitly exploit the sparse structure of the channel. To address this issue, works such as \cite{10959318, 10829527, cao2023efficient, cheng2019adaptive} adopted the sparse Bayesian learning framework \cite{fang2014pattern}, which incorporates the sparsity structure of the polar-domain channel, to further improve channel estimation accuracy. Codebook-based methods have demonstrated strong performance; however, their effectiveness heavily depends on the design of the codebook, which remains an open problem \cite{yu2023adaptive, zhang2023codebook}. In addition, the use of a codebook introduces storage overhead and increases system latency.

%\textbf{Deep learning methods}

Recently, deep learning (DL)-based approaches have gained attention due to their ability to learn from data, thereby reducing the need for precise knowledge of the channel structure. DL-based methods for channel estimation can be divided into two categories: model-free and model-based. Model-free DL approaches rely on a data-driven framework to directly map observations to wireless channels. In \cite{gao2024lightweight, chen2021hybrid, lei2023channel}, residual dense networks were proposed to refine rough channel estimates obtained from conventional approaches such as least-squares (LS). In \cite{jin2025near, arvinte2022mimo}, generative diffusion models were introduced for the CE task. The authors of \cite{jin2025near} proposed a two-stage approach: first, SOMP is applied to obtain a coarse channel estimate; second, a diffusion model is used to remove Gaussian noise from the coarse estimate.

Model-based DL methods, on the other hand, unroll iterative algorithms into a fixed number of layers, resulting in better accuracy and improved interpretability. For example, DL architectures built on the iterative shrinkage and thresholding algorithm were proposed in \cite{zhang2023near_, yang2025deep}, while those based on approximate message passing were presented in \cite{yu2023adaptive, jiang2024isac}. Although DL-based methods have proven effective for near-field channel estimation, they require a large amount of data for training. Furthermore, the performance of DL models is susceptible to degradation when deployed in scenarios that differ from the training environment.

%\textbf{Motivation}

Despite their effectiveness, the high computational and storage complexity of the aforementioned approaches limits their practicality in communication systems with stringent latency requirements. Moreover, most existing methods are designed for uniform linear arrays (ULAs), and extending them to uniform planar arrays (UPAs) is non-trivial. These challenges highlight the need for a low-complexity, scalable solution capable of handling more general array geometries.

In this work, we address these limitations by proposing an efficient channel estimation method based on the variational Bayes (VB) framework—a powerful statistical learning technique for approximating intractable posterior distributions using parameterized, tractable distributions. The VB framework has recently demonstrated strong performance in channel estimation and data detection across diverse MIMO systems. For example, \cite{nguyen2024variational} reports that VB achieves high joint channel estimation and data detection accuracy in low-bit ADC MIMO systems, whereas \cite{11044434, nassirpour2025variational} demonstrate its potential in cell-free massive MIMO and time-varying communication systems, respectively. For ELAAs-MIMO channel estimation, \cite{pisharody2024near} proposed a VB method that requires prior knowledge of both subarray-specific and shared sparsities. In particular, it considers a BS equipped with multiple antenna sub-arrays, each connected to a local processing unit (LPU). Each LPU estimates the channel of its corresponding sub-array and exchanges information with other LPUs to enable centralized channel estimation that exploits channel sparsity. However, this method may suffer from high computational overhead and still depends on prior knowledge of subarray-specific and shared sparsities. Moreover, it cannot estimate the positions of scatterers, which limits its potential for advanced transceiver design \cite{liu2024near}. Building on these foundations, our proposed approach eliminates the need for such prior information while preserving efficiency and scalability, and it further enables scatterer location estimation.

In summary, the main contributions of this work are as follows:
\begin{itemize}
    \item We develop a novel \underline{S}emi-\underline{G}ridless \underline{V}ariational \underline{B}ayes (SG-VB) for near-field channel estimation. By representing the near-field channel such that the frequency and distance components are separated into distinct latent variables, SG-VB becomes flexible for both ULAs and UPAs and achieves high estimation accuracy regardless of the array geometry. 
    
    \item We utilize a mixture of von Mises distributions to approximate frequencies and propose a coarse-to-fine grid search for scatterer distance estimation. 
    
    \item We derive closed-form variational distributions for all latent variables. These closed-form expressions provide two main benefits: (i) they enable efficient updates within the SG-VB algorithm, and (ii) they facilitate the estimation of scatterer positions, which can be leveraged for advanced transceiver design.
    \item We evaluate the performance of the proposed SG-VB method by comparing it with MUSIC, P-SOMP, P-SIGW, SBL, LS, and Oracle LS schemes in terms of normalized mean square error (NMSE) for channel reconstruction and distance estimation, as well as mean square error (MSE) for DoA estimation. Extensive simulation results verify the effectiveness of the proposed method across all metrics.
\end{itemize}

The rest of this paper is organized as follows. 
% Section II introduces the system and provides a new representation of ULA and UPA near-field channel modeling 
Section II introduces the system model and provides a new representation for ULA- and UPA-based near-field channel models. Section III reviews the VB framework and provides detailed derivations of SG-VB algorithms for both ULA and UPA. Section IV presents the simulation results, and Section V concludes the paper.

\underline{\textit {Notation}}: Bold lowercase letters denote vectors, and bold uppercase letters denote matrices. The symbols $\C$ and $\RR$ represent the sets of complex and real numbers, respectively. We use the symbol $\propto$ to denote ``proportional to''. 
The notation $\mathrm{diag}\{\mathbf{a}\}$ represents a diagonal matrix constructed from the vector $\mathbf{a}$. The $L_2$ norm and the absolute value are denoted by $\|\cdot\|$ and $|\cdot|$, respectively. The real and imaginary parts of a complex quantity are represented by $\mathfrak{R}
\{\cdot\}$ and $\mathfrak{I}\{\cdot\}$, respectively, where $\ji=\sqrt{-1}$. A circularly symmetric complex Gaussian (CSCG) distribution with mean $\boldsymbol{\eta}$ and covariance matrix $\mb{Z}$ is written as $\mathcal{CN}(\boldsymbol{\eta},\mb{Z})$. The transpose, complex conjugate, and Hermitian (conjugate transpose) operations are denoted by $(\cdot)^T$, $(\cdot)^*$, and $(\cdot)^H$, respectively. 

\section{System Model and Channel Modeling}
In this section, we first introduce the system model for the channel estimation framework. Next, we provide new representations for ULA and UPA near-field channel modeling. These new representations serve as foundation for SG-VB algorithms which will be discussed thoroughly in the next section. 
% \begin{tcolorbox}[colback=white!95!black, colframe=black!80!black, sharp corners, boxrule=0.8pt, width=\textwidth]
\subsection{System Model}
\begin{figure}[!t]
	\centering
\includegraphics[width=1.0\linewidth]{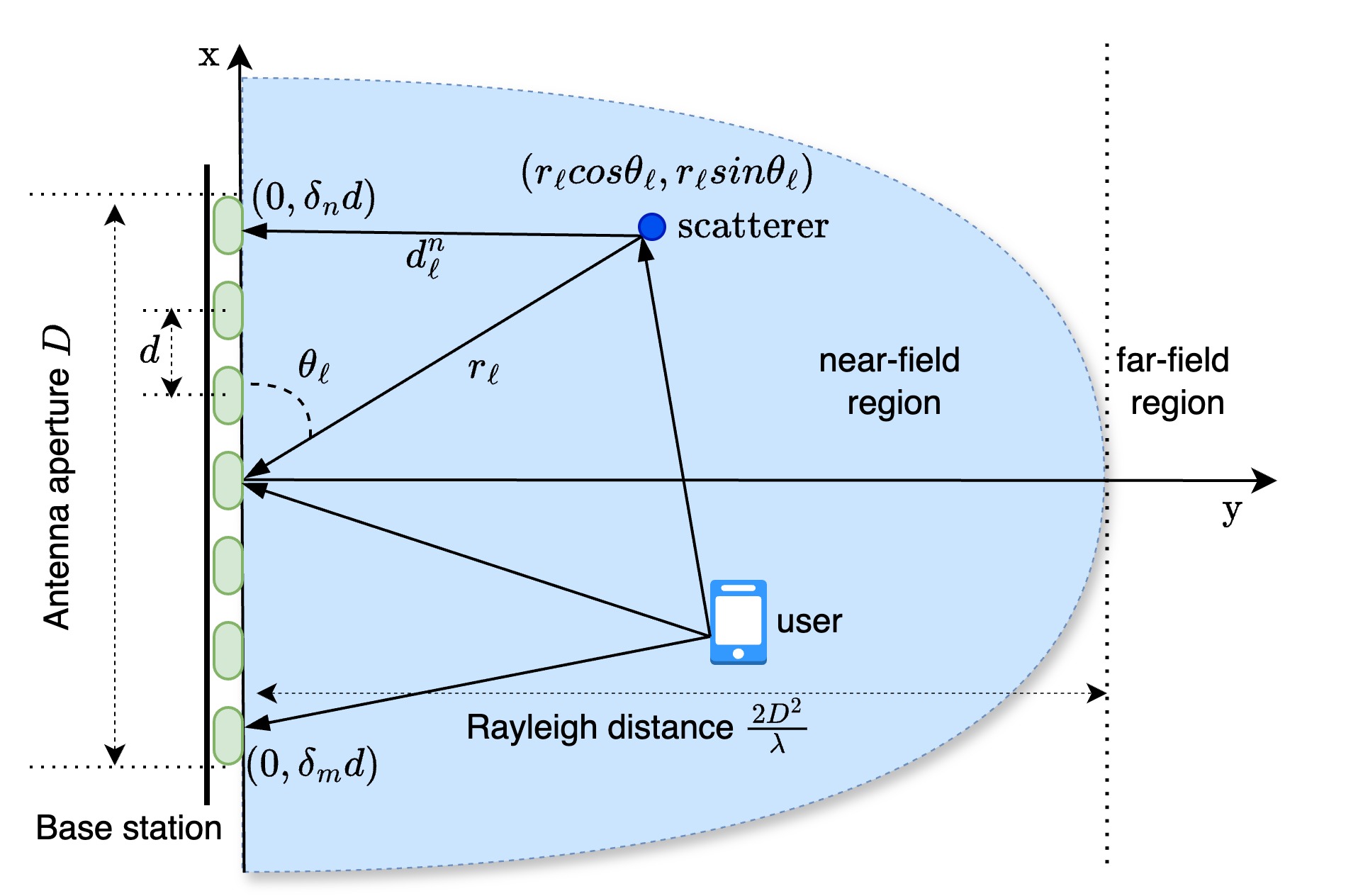}
	\caption{Illustration of the considered near-field system.}
	\label{fig:system_model}
\end{figure}

We consider the uplink of a near-field MIMO system, where the BS is equipped with a ULA or UPA with a total number of antennas $N$. The system model for the ULA is illustrated in Fig. \ref{fig:system_model}. The BS receives $L$ signal paths from multiple scatterers as well as a line-of-sight path from the user. In this work, we consider a fully digital system, as studied in \cite{gao2024lightweight, arvinte2022mimo, cao2023efficient, zhang2023near, lei2023channel}:
\begin{equation}
    \mathbf{y} = \mathbf{h} + \mathbf{n},
\end{equation}
where $\mathbf{y} \in \mathbb{C}^{N \times 1} $ is the received signal, $\mathbf{h} \in \mathbb{C}^{N \times 1}$ is the near-field channel, and $\mathbf{n} \sim \mathcal{CN}(\mathbf{0}, N_0 \mathbf{I}_N)$ presents i.i.d. complex Gaussian noise.% with noise variance $N_0$.

The main objective of this work is to estimate the near-field channel $\mathbf{h}$ from the noisy observation $\mathbf{y}$ with minimum ${\rm NMSE} \!=\!  10\log(\| {\mb{h}}-\hat{\mb{h}}\|^2/\|\mb{h}\|^2)$, with $\hat{\mb{h}}$ the estimated channel.

% \begin{tcolorbox}[colback=white!95!black, colframe=black!80!black, sharp corners, boxrule=0.8pt, width=\textwidth]
% the -$\ell$-th ray and the BS
\subsection{ULA Near-Field Channel Modeling}
Suppose the coordinates of the $\ell$-th scatter and the $n$-th antenna are $\big(r_\ell\cos\theta_\ell,r_\ell\sin\theta_\ell\big)$ and $(0,\delta_n)$, respectively, where $\delta_n = {(2n-N+1)d}/{2}$, $n = 0,\ldots,N-1$, and $d$ is the antenna spacing. In this case, the near-field channel between the $\ell$-th scatterer and the BS can be modeled as
\begin{equation}
\mathbf{h}_\ell = \alpha_\ell \, \mathbf{b}(\theta_\ell, r_\ell) \in \mathbb{C}^{ N\times 1}.
\end{equation}
Here, $\alpha_\ell$ stands for complex path gain and the steering vector $\mb{b}(r_\ell,\theta_\ell)$ corresponding to the $\ell$-th ray is defined as \cite{cui2022channel}
\begin{align}\mb{b}(r_\ell,\theta_\ell) = \big[\e^{\ji k(r_\ell - d^{(0)}_\ell)}, \ldots, \e^{\ji k(r_\ell - d^{(N-1)}_\ell)}\big]^T,
\end{align}
with $d_\ell^{(n)}$ being the distance between the $\ell$-th scatter and the $(n+1)$-th antenna, and $k = {2\pi}/{\lambda}$ denoting the wavenumber associated with the carrier frequency. Assuming the antenna array is deployed on the $x$-axis, the propagation distance is calculated as
$d_\ell^n = \sqrt{r_\ell^2 + \delta_n^2 - 2\delta_nr_\ell\sin\theta_\ell}$, which is further approximated using the Fresnel approximation \cite{selvan2017fraunhofer} as
\begin{align}
    d_\ell^n 
    %&= r_\ell \sqrt{1 + \frac{\delta_n^2}{r_\ell^2} - \frac{2\delta_n\sin\theta_\ell}{r_\ell}} \nonumber  \\
    \approx r_\ell\left(1 + \frac{\delta_n^2}{2r_\ell^2} - \frac{\delta_n\sin\theta_\ell}{r_\ell}\right) 
    = r_\ell + \frac{\delta_n^2}{2r_\ell} - \delta_n\sin\theta_\ell.
\end{align}
We can now write the $n$-th element of $\mb{b}(r_\ell,\theta_\ell)$ as
\begin{align}
    \big[\mb{b}(r_\ell,\theta_\ell)\big]_{n} &= \e^{\ji k\big(\delta_n\sin\theta_\ell -\frac{\delta_n^2}{2r_\ell}\big)} \nonumber \\
    &= \e^{-\ji\frac{N-1}{2}kd\big(\sin\theta_\ell+\frac{(N-1)d}{4r_\ell}\big)} \e^{\ji nkd\sin\theta_\ell} \nonumber \\
    &\quad\times\e^{\ji((N-1)n-n^2)\frac{kd^2}{2r_\ell}}.
\end{align}
Defining
$\nu_\ell = \alpha_\ell \e^{-\ji\frac{N-1}{2}kd\big(\sin\theta_\ell+\frac{(N-1)d}{4r_\ell}\big)}$, $\omega_\ell = kd\sin\theta_\ell$, and $s_\ell = \frac{kd^2}{2r_\ell}$, the received signal at the $n$-th antenna can be expressed as
\begin{align}
    y_n = \sum_{\ell=1}^L \nu_\ell \e^{\ji n\omega_\ell} \e^{\ji n(N-1-n)s_\ell} + z_n,
\end{align}
while the signal at the array can be represented in vector form as
\begin{align} \label{system-model_ula}
    \mb{y} = \sum_{\ell=1}^L \nu_\ell \mb{a}(\omega_\ell) \odot \mb{c}(s_\ell) + \mb{n},
\end{align}
where $\mb{a}(\omega_\ell) \triangleq \big[1,\e^{\ji\omega_\ell},\ldots,\e^{\ji (N-1)\omega_\ell}\big]^T \in \mathbb{C}^{N}$, $\mb{c}(s_\ell) \triangleq \big[1,\e^{\ji (N-2) s_\ell},\e^{\ji 2(N-3)s_\ell},\ldots,\e^{\ji 2(N-3)s_\ell},\e^{\ji (N-2)s_\ell}, 1\big]^T \in \mathbb{C}^{N}$, and $\odot$ denotes the element-wise product of the two vectors.
\begin{Remark}\label{remark:ula}
As $r_\ell\rightarrow \infty$ in far-field transmission, $s_\ell$ effectively vanishes, i.e., $s_\ell\rightarrow 0$ and the array response vector $\mb{c}(s_\ell)$ approaches an all-$1$ vector. As a result, the system model in \eqref{system-model_ula} becomes the model that is widely known for transmission from the far field.
\end{Remark}

\subsection{UPA Near-Field Channel Modeling}
Let \(N_{\mathrm{H}}\) and \(N_{\mathrm{V}}\) denote the number of antennas per row and per column of the UPA, respectively. The total number of antennas is \(N = N_{\mathrm{H}} N_{\mathrm{V}}\). We consider identical and uniform spacing \(\Delta\) between adjacent antennas in both vertical and horizontal directions. The antennas are sequentially indexed row by row, with the index parameter \(i \in \{1, \ldots, N\}\). Assuming the antenna array is deployed on the $yz$-plane, the position of the \(i\)-th antenna relative to the origin is given by the vector
\[
\mathbf{u}_i = [0, \; m_i \Delta, \; n_i \Delta]^{\mathrm{T}},
\]
where
\[
m_i = \mathrm{mod}(i-1, N_{\mathrm{H}}) \quad \text{and} \quad n_i = \left\lfloor \frac{i-1}{N_{\mathrm{H}}} \right\rfloor
\]
represent the horizontal and vertical indices, respectively. Here, \(\mathrm{mod}(\cdot, \cdot)\) is the modulus operation, while \(\lfloor \cdot \rfloor\) is the truncation (floor) operation. The near-field channel between the $\ell$-th scatterer and the BS can be modeled as
% \textcolor{blue}{Van: "the $k$-th user and the BS": how many users here? or just the scatterers?}
\begin{equation}
\mathbf{h}_\ell = \alpha_\ell \, \mathbf{b}(r_\ell,\theta_\ell, \varphi_\ell) \in \mathbb{C}^{ N \times 1},
\end{equation} with
\begin{align}\mb{b}(r_\ell,\theta_\ell, \varphi_\ell) = \big[\e^{\ji k(r_\ell - d^{(1)}_\ell)}, \ldots, \e^{\ji k(r_\ell - d^{(N)}_\ell)}\big]^T,
\end{align} where the distance \(d_\ell^{i}\) from the BS antenna \(i\) to the scatterer is given by
\begin{align}
d_\ell^{i}
&= \left[
\bigl(r_\ell\cos\theta\cos\varphi\bigr)^2
+ \bigl(r_\ell\cos\theta\sin\varphi - m_i\Delta\bigr)^2 \right. \nonumber \\
&\quad\left. + \bigl(r_\ell\sin\theta - n_i\Delta\bigr)^2
\right]^{\frac{1}{2}} \nonumber \\
&= r_\ell\left[
1 - \frac{2\Delta}{r_\ell}\bigl(m_i\cos\theta\sin\varphi + n_i\sin\theta\bigr)
+ \frac{\Delta^2}{r_\ell^2}(m_i^2 + n_i^2)
\right]^{\frac{1}{2}}.
\end{align}
The distance $d_\ell^{i}$ can be further approximated \cite{selvan2017fraunhofer} as
\begin{equation}
d_\ell^{i} \approx r_\ell - \Delta \left( m_i \cos(\theta) \sin(\varphi) + n_i \sin(\theta) \right) + \Delta^2 \left( \frac{m_i^2 + n_i^2}{2r_\ell} \right).
\end{equation}
Hence, the $i$-th element of the steering vector $\mathbf{b}(r_\ell,\theta_\ell, \varphi_\ell)$ is given by

\begin{align}
\mathbf{b}_i(r_\ell,\theta_\ell, \varphi_\ell) 
&= \exp\!\left( \ji \frac{2\pi}{\lambda} \Delta \, m_i \cos\theta \sin\varphi \right) \\
&\quad \times \exp\!\left( \ji \frac{2\pi}{\lambda} \Delta \, n_i \sin\theta \right) \nonumber \\
&\quad \times \exp\!\left( -\ji \frac{2\pi}{\lambda} \cdot \frac{\Delta^2}{2r_\ell} 
(m_i^2 + n_i^2) \right). \nonumber
\end{align}
As a result, the UPA array response vector can be represented as
\begin{align}
\mathbf{b}(r_\ell,\theta_\ell, \varphi_\ell) 
&=\mathbf{a}(\omega_\ell) \odot \mathbf{c}(\psi_\ell) \odot \mathbf{d}(s_\ell),
\end{align}
where $\mb{a}(\omega_\ell) \triangleq \big[e^{\ji \omega m_1},\ldots,e^{\ji \omega m_N}\big]^T \in \mathbb{C}^{N}$, $\mb{c}(\psi_\ell) \triangleq \big[e^{\ji \psi m_1},\ldots,e^{\ji \psi m_N}\big]^T \in \mathbb{C}^{N}, \mb{d}(s_\ell) \triangleq \big[e^{-\ji s \left(m_1^2+n_1^2 \right)},\ldots,e^{-\ji s \left(m_M^2+n_M^2 \right)}\big]^T \in \mathbb{C}^{N}$
and 
\[
\omega = \frac{2\pi}{\lambda} \Delta \cos\theta \sin\varphi,
\quad
\psi = \frac{2\pi}{\lambda} \Delta \sin\theta,
\quad
s = \frac{2\pi \Delta^2}{\lambda \cdot 2r_\ell}.
\]

Finally, the received signal can be expressed as
\begin{align} \label{system-model}
    \mb{y} = \sum_{\ell=1}^L \alpha_\ell \mathbf{a}(\omega_\ell) \odot \mathbf{c}(\psi_\ell) \odot \mathbf{d}(s_\ell) + \mb{n}.
\end{align}

\section{Variational Inference of Near-Field Parameters}
Building on the new representations of near-field channels introduced in the previous section, we develop the SG-VB algorithm for near-field CE. Specifically, we first review the background of the VB method, and then derive the closed-form updates for the DoA and distance components for both ULA and UPA.

\subsection{Background on Variational Bayes Inference}
 Denote the set of all observed variables by $\mb{y}$ and the set of $m$ latent variables and parameters by $\mb{h}$. The probabilistic model specifies the joint distribution
$p(\mb{y},\mb{h})$. The estimation problem aims to find the posterior $p(\mb{h}|\mb{y})$, which is often computationally intractable. To overcome this issue, variational inference aims to identify a distribution \( q(\mb{h}) \), parameterized within a family of densities \( \mc{Q} \), that closely approximates the true posterior \( p(\mb{h}|\mb{y}) \) by minimizing the Kullback-Leibler (KL) divergence \cite{bishop2006pattern, wainwright2008graphical}
\begin{eqnarray}
    q^{\star}(\mathbf{h}) 
    &=& \arg\min_{q(\mathbf{h}) \in \mathcal{Q}} \;
    \mathrm{KL}\!\left( q(\mathbf{h}) \,\|\, p(\mathbf{h} \mid \mathbf{y}) \right).
\end{eqnarray}
This is equal to maximize the evidence lower bound ($\mr{ELBO}$) defined as
	\begin{eqnarray}
		\mr{ELBO}(q) =  \E_{q(\mb{h})} \big[\ln p(\mb{h},\mb{y})\big] - \E_{q(\mb{h})} \big[\ln q(\mb{h}) \big].
	\end{eqnarray}
    The maximum of $\mr{ELBO}(q)$ occurs when $q(\mb{h}) = p(\mb{h}|\mb{y})$. In addition, VB relies on the \emph{mean-field variational family} \cite{bishop2006pattern}
	\begin{eqnarray}\label{mean-field}
		q(\mb{h}) = \prod_{i=1}^m q_i(h_i),
	\end{eqnarray}
	in which the latent variables are assumed to be mutually independent, each associated with a separate factor in the variational distribution. The optimal solution of the variational density $q_i(h_i)$ is \cite{bishop2006pattern}
	\begin{eqnarray}
		q_i^{\star}(h_i) \propto \mr{exp}\left\{\big\langle{\ln p (\mb{y}|\mb{h}) + \ln p(\mb{h})\big\rangle}_{-h_i}\right\}.
	\end{eqnarray}
	Here, $\lr{\cdot}_{-h_i}$ denotes the expectation with respect to all latent variables except $h_i$ using the currently fixed variational density $q_{-i}(\mb{h}_{-i}) = \prod_{j\neq i} q_{j}(h_{j})$. By sequentially updating \( q_i^{\star}(h_i) \) over all \( j \), the objective function \( \mathrm{ELBO}(q) \) increases monotonically. This iterative scheme forms the foundation of the \emph{Coordinate Ascent Variational Inference (CAVI)} algorithm, which guarantees convergence to at least a local optimum of \( \mathrm{ELBO}(q) \) \cite{bishop2006pattern, wainwright2008graphical}.

\subsection{Proposed SG-VB for ULA Channel Estimation}
We define $\gamma \triangleq 1/N_0$ as the precision, treated as an unknown random variable to be estimated within the VB framework.
The objective is to infer the distributions of the DoA, $\bs{\theta} = [\theta_1,\ldots,\theta_L]^T$, and the distances, $\mb{r} = [r_1,\ldots,r_L]^T$, using the auxiliary variables $\bs{\omega} = [\omega_1,\ldots,\omega_L]^T$ and $\mb{s} = [s_1,\ldots,s_L]^T$, respectively, given the observation $\mb{y}$.
To achieve this, we adopt the mean-field variational distribution $q(\bs{\nu},\bs{\omega},\mb{s},\gamma)$, such that
\begin{align} \label{eq:mean-field}
p(\bs{\nu},\bs{\omega},\mb{s},\bs{\beta},\gamma|\mb{y})&\approx q(\bs{\nu},\bs{\omega},\mb{s},\bs{\beta},\gamma) \nonumber \\
&= q(\bs{\nu})q(\bs{\omega})q(\bs{s})q(\bs{\beta})q(\gamma) \nonumber\\
&= q(\gamma) \Bigg[\prod_{\ell=1}^L q(\nu_{\ell})q(\omega_{\ell})  q(s_{\ell}) q(\beta_{\ell})\Bigg].
\end{align}

Based on the VB framework, to obtain the optimal solution of the variational densities in \eqref{eq:mean-field}, we need the joint distribution $p(\mb{y},\bs{\nu},\bs{\omega},\mb{s},\bs{\beta},\gamma)$ of the observed variables $\mb{y}$ and the latent variables $\bs{\nu},\bs{\omega},\mb{s},\bs{\beta},\gamma$, which can be factorized as
\begin{align} \label{eq:jointPDF}
p(\mb{y},\bs{\nu},\bs{\omega},\mb{s},\bs{\beta},\gamma) = p(\mb{y}|\bs{\nu},\bs{\omega},\mb{s},\gamma) p(\bs{\nu}|\bs{\beta})p(\bs{\omega})p(\mb{s})p(\bs{\beta})p(\gamma).
\end{align}

Since near-field channels are inherently sparse, a Gamma-Gaussian prior is adopted for the path gain to enforce sparsity in the estimated propagation paths, enabling closed-form Bayesian inference. Thus, $\nu_{\ell}$ is assumed to be drawn from a zero-mean Gaussian distribution, i.e., $p(\nu_{\ell}|\beta_{\ell}) = \mc{CN}\big(\nu_{\ell};0,\beta_{\ell}^{-1}\big)$, where the variance $\beta_{\ell}$ follows Gamma distribution $p(\beta_{\ell}) = \Gamma(a_{\beta},b_{\beta})$.

\emph{1) Update $\omega_\ell$:} Taking the expectation of the conditional \eqref{eq:jointPDF} w.r.t. all latent variables except $\omega_{\ell}$, the variational distribution $q(\omega_{\ell})$ can be obtained as
\begin{align}
    q(\omega_\ell) &\propto p(\omega_\ell) \exp\big\{\lr{\ln p(\mb{y}|\bs{\nu},\bs{\omega},\mb{s},\gamma)}_{-\omega_\ell}\big\}\nonumber\\
    &\propto p(\omega_\ell) \exp\Bigg\{-\bblr{\gamma\bigg\|\mb{y} - \sum_{\ell=1}^L \nu_\ell \mb{a}(\omega_\ell) \odot \mb{c}(s_\ell)\bigg\|^2}_{-\omega_\ell}\Bigg\} \nonumber \\
    &\propto p(\omega_\ell) \exp\Bigg\{-\bblr{\gamma\bigg\|\mb{y} - \sum_{k\neq \ell}^L \nu_k \mb{a}(\omega_k) \odot \mb{c}(s_k) \nonumber\\
    &\qquad\qquad\qquad\qquad - \nu_\ell \mb{a}(\omega_\ell) \odot \mb{c}(s_\ell)\bigg\|^2}_{-\omega_\ell}\Bigg\} \nonumber \\
 %   &\propto p(\omega_\ell) \exp\bigg\{-\sum_{n=1}^N\bblr{\gamma\bigg\|y_n-\sum_{k\neq \ell}^L\nu_k\e^{\ji n\omega_k}\e^{\ji n(N-1-n)s_k} - \nu_\ell \e^{\ji n\omega_\ell}\e^{\ji n(N-1-n)s_\ell}\bigg\|^2}\bigg\} \nonumber\\
 %   &\propto p(\omega_\ell) \exp\bigg\{\sum_{n=1}^N\Re\big\{\eta_{n,\ell}^*\e^{\ji n\omega_\ell}\big\} \bigg\} \nonumber \\
    &\propto p(\omega_\ell) \exp\big\{\Re\big\{\bs{\eta}_{\ell}^H \mb{a}(\omega_\ell)\big\} \big\},
\end{align}
where 
\begin{align}
    \bs{\eta}_{\ell} &= 2\hat{\gamma}\hat{\nu}_\ell^* \blr{\mb{c}^*(s_\ell)} \odot \bigg(\mb{y} - \sum_{k\neq \ell}^L \hat{\nu}_k \blr{\mb{a}(\omega_k)}\odot\blr{\mb{c}(s_k)}\bigg) \nonumber \\
    &=2\hat{\gamma}\hat{\nu}_\ell^* \hat{\mb{c}}^*_\ell \odot (\bs{\varepsilon} + \nu_\ell \hat{\mb{a}}_\ell\odot \hat{\mb{c}}_\ell).
\end{align}
Here, for notational convenience, we denote $\hat{\mb{c}}_{\ell} = \blr{\mb{c}(s_\ell)}$, $\hat{\mb{a}}_{\ell} = \blr{\mb{a}(\omega_\ell)}$ as the \emph{current} variational mean of $\mb{a}(\omega_\ell)$, and $\bs{\varepsilon} = \mb{y} - \sum_{\ell=1}^L \hat{\nu}_\ell \hat{\mb{a}}_\ell \odot \hat{\mb{c}}_\ell$ as a residual term using the current estimates of $\nu_\ell$, $\omega_\ell$, and $s_\ell$, $\forall \ell=1,\ldots,L$. In line with the principles of variational line spectral estimation \cite{badiu2017variational}, $q(\omega_{\ell})$ is approximated by a von Mises distribution using Heuristic algorithm 2, with mean $\hat{\omega}_\ell$ and concentration parameter $\kappa_{\omega_\ell}$, from which the expected steering vector is derived as $\hat{\mb{a}}(\omega_\ell) =\mb{A}(\kappa_{\omega_\ell})\mb{a}(\hat{\omega}_\ell)$, where 
\begin{align}
     \mb{A}(\kappa_{\omega_\ell}) = \mr{diag}\left(1,\frac{I_1(\kappa_{\omega_\ell})}{I_0(\kappa_{\omega_\ell})},\ldots,\frac{I_{N-1}(\kappa_{\omega_\ell})}{I_0(\kappa_{\omega_\ell})} \right).
\end{align}
The von Mises parameterization allows closed-form variational updates while eliminating the spectral discretization artifacts that arise in grid-based compressed sensing methods. Additional details on the von Mises distribution are provided in Appendix~\ref{appendix:vm}.

% Similarly, should the variational distribution of $s_\ell$ be von Mises with mean $\hat{s}_\ell$ and concentration $\kappa_{s_\ell}$, one has $\hat{\mb{c}}(s_\ell) = \mb{C}(\kappa_{s_\ell})\mb{c}(\hat{s}_\ell).$
% \begin{align}
%     \mb{C}(\kappa_{s_\ell})= \mr{diag}\left(1,\frac{I_{N-2}(\kappa_{s_\ell})}{I_0(\kappa_{s_\ell})},\frac{I_{2(N-3)}(\kappa_{s_\ell})}{I_0(\kappa_{s_\ell})},\ldots,\frac{I_{2(N-3)}(\kappa_{s_\ell})}{I_0(\kappa_{s_\ell})},\frac{I_{N-2}(\kappa_{s_\ell})}{I_0(\kappa_{s_\ell})}, 1 \right).
% \end{align}
% \begin{align}
%     \mb{C}(\kappa_{s_\ell})= \mr{diag}\left(1,\frac{I_{N-2}(\kappa_{s_\ell})}{I_0(\kappa_{s_\ell})},\ldots,\frac{I_{N-2}(\kappa_{s_\ell})}{I_0(\kappa_{s_\ell})}, 1 \right).
% \end{align}
%Thus $q(\omega_\ell)$ can be approximated as a von Mises distribution function.

%We note that $\|{\mb{a}}(\omega_{\ell})\|^2 = \sum_{n=0}^{N-1}|\e^{jn\omega_\ell}|^2 = N$, $\|{\mb{c}}(\omega_{\ell})\|^2 = N$, and $\|{\mb{a}}(\omega_{\ell})\odot {\mb{c}}(\omega_{\ell})\|^2 = N$.

\emph{2) Update $s_\ell$:} Taking the expectation of the conditional \eqref{eq:jointPDF} w.r.t. all latent variables except $s_{\ell}$, the variational distribution $q(s_{\ell})$ can be derived as
\begingroup
\allowdisplaybreaks
\begin{align} \label{eq:q_s_l}
    q(s_\ell) &\propto p(s_\ell) \exp\big\{\lr{\ln p(\mb{y}|\bs{\nu},\bs{\omega},\mb{s},\gamma)}_{-s_\ell}\big\}\nonumber\\
    &\propto p(\omega_\ell) \exp\Bigg\{-\bblr{\gamma\bigg\|\mb{y} - \sum_{k\neq \ell}^L \nu_k \mb{a}(\omega_k) \odot \mb{c}(s_k) \nonumber\\
    &\qquad \qquad\qquad -\nu_\ell \mb{a}(\omega_\ell) \odot \mb{c}(s_\ell)\bigg\|^2}_{-s_\ell}\Bigg\} \nonumber \\
    &\propto p(s_\ell) \exp\big\{\Re\big\{\bs{\zeta}_\ell^H \mb{c}(s_\ell)\big\}\big\},
\end{align}
\endgroup
where 
\begin{align}
    \bs{\zeta}_\ell &= 2\hat{\gamma}\hat{\nu}_\ell^* \blr{\mb{a}^*(\omega_\ell)} \odot \bigg(\mb{y} - \sum_{k\neq \ell}^L \hat{\nu}_k \blr{\mb{a}(\omega_k)}\odot\blr{\mb{c}(s_k)}\bigg) \nonumber \\
    &=2\hat{\gamma}\hat{\nu}_\ell^* \hat{\mb{a}}^*_\ell \odot (\bs{\varepsilon} + \nu_\ell \hat{\mb{a}}_\ell\odot \hat{\mb{c}}_\ell).
\end{align}
Note that $s_\ell$ has to be positive; therefore, if the variational mean $\langle s_\ell \rangle$ is negative, we set the estimate $\hat{s}_\ell$ to $0$ as discussed in Remark~\ref{remark:ula}, which corresponds to user-$\ell$ being in the far-field. To find the optimal $s_\ell$, we maximize the log-likelihood of $s_\ell$ given the residual correlation:
  \begin{align}
    [\mc{L}_\ell]_k = \Re\{{\bs{\zeta}}_\ell^H \mb{c}(s_k)\},
\end{align}
where $s_k \in \mc{S}_{\rm grid}$ for $k = 1,\ldots,K$, and $K$ denotes the number of coarse grid points in the set $\mc{S}_{\rm grid}$.

To this end, we propose a coarse-to-fine grid search strategy described as follows.
First, $s_\ell$ is estimated over a coarse grid using the maximum likelihood criterion:
\begin{align}
    \hat{s}_\ell = \arg\max_{k} \bs{\mc{L}}_\ell,
\end{align} 
with $s_\ell \in \{s_k\}_{k=1}^K$.
Next, the first- and second-order derivatives of $[\mc{L}_\ell]_k$ with respect to $s_\ell$ are given by
\begin{align}
    \Delta[\mc{L}_\ell]_k &= -\Im\{{\bs{\zeta}}_\ell^H (\mb{m}\odot\mb{c}(s_k))\},\\
    \Delta^2 [\mc{L}_\ell]_k &= -\Re\{{\bs{\zeta}}_\ell^H (\mb{m}^2\odot\mb{c}(s_k))\},
\end{align}
where the index vector is defined as $\mb{m} \triangleq [0,\ldots,n(N-1-n),\ldots,0]^T \in \mathbb{C}^N$.
These derivatives are then used to iteratively refine the estimate of $s_\ell$ via the Newton–Raphson update:
\begin{align}
s_\ell^{t+1} = s_\ell^t - \epsilon \frac{\Delta\mc{L}(s_\ell^t)}{\Delta^2\mc{L}(s_\ell^t)},
\end{align}
where the step size $\epsilon$ is typically set to $0.01$. A summary of the proposed GridSearch procedure is provided in Algorithm~\ref{algo-1}.
\begin{algorithm}[t]
\small
	\caption{: Proposed GridSearch Algorithm for Estimation of $s_\ell$}
	\begin{algorithmic}[1] 
		\STATE \textbf{Input:}
		${\bs{\zeta}}_\ell^H \mb{c}(s_k)$ and $\mc{S}_{\rm{grid}}$
        \STATE \textbf{Output:}	$s_\ell$.
        \STATE Compute all $[\mc{L}_\ell]_k \gets \Re({\bs{\zeta}}_\ell^H \mb{c}(s_k))$, $\forall k \in \mc{S}_{grid}$
        \STATE $\hat{s}_\ell = \arg\max\limits_{k} \bs{\mc{L}}_\ell$
        \WHILE{$\xi>0$}
        \STATE $s_\ell^{t+1} \gets s_\ell^t - \epsilon\frac{\Delta\mc{L}(s_\ell^t)}{\Delta^2\mc{L}(s_\ell^t)},$
        \STATE $\xi \gets |\mc{L}_\ell(s_\ell^{t+1}) - \mc{L}_\ell(s_\ell^{t})|$
        \ENDWHILE
	\end{algorithmic} 
    \label{algo-1}
\end{algorithm}

\emph{3) Update $\nu_\ell$:} The variational distribution $q(\nu_\ell)$ can be derived as
\begingroup
\allowdisplaybreaks
\begin{align}
    q(\nu_\ell) &\propto \exp\big\{\lr{\ln p(\mb{y}|\bs{\nu},\bs{\omega},\mb{s},\gamma) + \ln p(\nu_\ell|\beta_\ell)}_{-\nu_\ell}\big\}\\
    &\propto \mc{CN}\bigg\{-\hat{\gamma}\big(N|\nu_\ell|^2 - 2\,\Re\bigg\{\nu_\ell \bigg(\mb{y} - \sum_{k\neq \ell}^L \hat{\nu}_k \hat{\mb{a}}_\ell \odot \hat{\mb{c}}_\ell \bigg)^H \nonumber\\
&\qquad\qquad\times\big(\hat{\mb{a}}_\ell\odot\hat{\mb{c}}_\ell\big)\bigg\} - \lr{\beta_\ell} |\nu_\ell|^2\bigg\} \nonumber \\
    &\propto \mc{CN}\big(\nu_\ell; z_\ell, (N\hat{\gamma})^{-1}\big) \mc{CN}\big(\nu_\ell;0,\hat{\beta}_\ell^{-1}\big) \nonumber \\
    &\propto \mc{CN}\big(\nu_\ell; \hat{\nu}_\ell, \tau_{\nu_\ell}\big),
    \label{eq:q_nu_update}
\end{align}
\endgroup
where
%\begin{align}
    $z_\ell \triangleq {(\hat{\nu}_\ell\big\|\hat{\mb{a}}_\ell\odot\hat{\mb{c}}_\ell\big\|^2 +  \big(\hat{\mb{a}}_\ell\odot\hat{\mb{c}}_\ell\big)^H\bs{\varepsilon})}/{N}$, 
%\begin{align}
    ${\tau_{\nu_\ell} = {(N\hat{\gamma} + \hat{\beta}_\ell)^{-1}}}$, and $
    \hat{\nu}_\ell = {z_\ell N\hat{\gamma}}{(N\hat{\gamma} + \hat{\beta}_\ell)^{-1}}.$
Eq.~(\ref{eq:q_nu_update}) is obtained by using the property of the product of two Gaussian densities.
\footnote{Property of the product of two Gaussian densities\label{ft:gaussian_property} 
\begin{align*} 
&\mc{CN}(\mb{x};\mb{a},\mb{A})\,\mc{CN}(\mb{x};\mb{b},\mb{B})  = \mc{CN}(\mb{0};\mb{a}-\mb{b},\mb{A}+\mb{B}) \\
&\quad\times\mc{CN}\big(\mb{x};(\mb{A}^{-1}+\mb{B}^{-1})^{-1}(\mb{A}^{-1}\mb{a} + \mb{B}^{-1}\mb{b}),(\mb{A}^{-1}+\mb{B}^{-1})^{-1}\big).
\end{align*}} 

\emph{4) Update $\beta_\ell$:} The variational distribution $q(\beta_\ell)$ can be obtained as 
\begin{align}
    q(\beta_\ell) &\propto p(\beta_\ell) \exp\{\lr{p(\nu_\ell|\beta_\ell)}\}.
\end{align}
Thus, $q(\beta_\ell)$ is a Gamma distribution with mean ${\hat{\beta}_\ell = (|\hat{\nu_\ell}|^2 + \tau_{\nu_\ell})^{-1}}$.

\emph{5) Update $\gamma$:} The variational distribution $q(\gamma)$ can be derived as
\begin{align*}
    q(\gamma) &\propto p(\gamma) \exp\big\{\lr{\ln p(\mb{y}|\bs{\nu},\bs{\omega},\mb{s},\gamma)}_{-\gamma}\big\}\\
    &\propto p(\gamma) \exp\Bigg\{N\ln\gamma-\gamma\bblr{\bigg\|\mb{y} \!-\! \sum_{\ell=1}^L \nu_\ell \mb{a}(\omega_\ell) \!\odot\! \mb{c}(s_\ell)\bigg\|^2}\Bigg\}.
\end{align*}

Thus, $q(\gamma)$ is a Gamma distribution with mean 
\begin{align}\label{eq:update_gamma}
 \hat{\gamma} = \frac{a_{\gamma} + N}{b_\gamma + \blr{\big\|\mb{y} - \sum_{\ell=1}^L \nu_\ell \mb{a}(\omega_\ell) \odot \mb{c}(s_\ell)\big\|^2}}, 
\end{align}
where
\begin{align}\label{eq:update_gamma}
   \bblr{\bigg\|\mb{y} - \sum_{\ell=1}^L \nu_\ell \mb{a}&(\omega_\ell) \odot \mb{c}(s_\ell)\bigg\|^2} = \|\bs{\varepsilon}\|^2 + \sum_{\ell=1}^L \tau_{\nu_\ell}\|\hat{\mb{a}}_\ell\odot\hat{\mb{c}}_\ell \|^2 \nonumber\\
    & + \sum_{\ell=1}^L \big(|\hat{\nu}_\ell|^2 +\tau_{\nu_\ell}\big)\big(N - \|\hat{\mb{a}}_\ell\odot\hat{\mb{c}}_\ell \|^2\big).
\end{align}

The expansion of (\ref{eq:update_gamma}) comes from Theorem 1 in \cite{nguyen2022variational}.
By iteratively updating $\omega_\ell, s_\ell, \nu_\ell, \gamma$, we obtain the SG-VB algorithm for near-field channel estimation problem, which is summarized in Algorithm \ref{algo-2}.
\begin{algorithm}[t]
\small
	\caption{: SG-VB Algorithm for ULA Near-Field Channel Estimation}
	\begin{algorithmic}[1] 
		\STATE \textbf{Input:}
		$\textbf{y}$
        \STATE \textbf{Output:}	$\hat{\bm{\omega}}$, $\hat{\bm{\nu}}$, $\hat{\mb{s}}$, and $\hat{\mb{h}}$
		\STATE Initialize $\hat{\mb{h}} = \mb{y}$, $\bs{\varepsilon}^1 = \mb{y} - \sum_{\ell=1}^L \hat{\nu}_\ell \hat{\mb{a}}_\ell \odot \hat{\mb{c}}_\ell$, and $\hat{\bs{\nu}}^1 = 0$.
        \FOR{$t = 1,2,\ldots$ } 
        \FOR{$\ell = 1,\ldots,L$} 
        % update freq omega
        \STATE $\bs{\eta}_\ell = 2\hat{\gamma}^t(\hat{\nu}_\ell^t)^* (\hat{\mb{c}}_\ell^t)^* \odot (\bs{\varepsilon} + \nu_\ell^t \hat{\mb{a}}_\ell^t\odot \hat{\mb{c}}_\ell^t)$
        \STATE $[\hat{\omega}_{\ell}^t,\, \hat{\kappa}_{s_\ell}^t]  = {\rm Heuristic2}(\bs{\eta_{\ell}})$
        \STATE $\hat{\mb{a}}_\ell^{t+1} \leftarrow \mb{A}(\hat{\kappa}_{s_\ell}^t) {\mb{a}}_\ell( \hat{\omega}_{\ell}^t)$
        \STATE $\bs{\varepsilon} \leftarrow \bs{\varepsilon} +  \hat{\nu}_\ell^t (\hat{\mb{a}}_\ell^t -\hat{\mb{a}}_\ell^{t+1}) \odot \hat{\mb{c}}_\ell^t$ 
        %% update distance s_l
        \STATE $\bs{\zeta}_\ell \leftarrow 2\hat{\gamma}^t(\hat{\nu}_\ell^t)^* (\hat{\mb{a}}^{t+1}_\ell)^* \odot (\bs{\varepsilon} + \nu_\ell^t \hat{\mb{a}}_\ell^{t+1}\odot \hat{\mb{c}}_\ell^t)$
        \STATE $\hat{s}_{\ell}^{t+1}  = {\rm GridSearch}(\bs{\zeta}_{\ell})$ 
        \STATE $\hat{\mb{c}}_\ell^{t+1} \leftarrow {\mb{c}}_\ell( \hat{s}_{\ell}^{t+1})$
        \STATE $\bs{\varepsilon} \leftarrow \bs{\varepsilon} + \hat{\nu}_\ell^t \hat{\mb{a}}_\ell^{t+1} \odot (\hat{\mb{c}}_\ell^t -\hat{\mb{c}}_\ell^{t+1})$
        % update \nu
        \STATE $z_\ell = \Big[ \big(\hat{\mb{a}}_\ell^{t+1}\odot\hat{\mb{c}}_\ell^{t+1}\big)^H\bs{\varepsilon}+ \hat{\nu}_\ell^t\big\|\hat{\mb{a}}_\ell^{t+1}\odot\hat{\mb{c}}_\ell^{t+1}\big\|^2\Big]\big/N$
        \STATE  $\tau_{\nu_{\ell}}^{t+1}  \leftarrow (\hat{\beta}_\ell^t + N\hat{\gamma}^t)^{-1}$
        \STATE $\hat{\nu}_{\ell}^{t+1} \leftarrow {z_\ell N\hat{\gamma}^t}\tau_{\nu_{\ell}}^{t+1}$
        \STATE $\bs{\varepsilon} \leftarrow \bs{\varepsilon} +  (\hat{\nu}_\ell^t-\hat{\nu}_\ell^{t+1}) \hat{\mb{a}}_\ell^{t+1} \odot \hat{\mb{c}}_\ell^{t+1}$
        \ENDFOR
        \STATE $\bs{\beta}^{t+1} \leftarrow {(a_\beta + 1)}/{(b_\beta + \|\hat{\bs{\nu}}^{t+1}\|^2 + \bs{\tau}_{\bs{\nu}_{i}}^{t+1})}$ 
        \STATE $\bs{\varepsilon} \leftarrow \mb{y} - \sum_{\ell=1}^L \hat{\nu}_\ell^{t+1} \hat{\mb{a}}_\ell^{t+1} \odot \hat{\mb{c}}_\ell^{t+1}$
        \STATE $\gamma^{t+1} \leftarrow (a_\gamma + N)/\big(b_\gamma + \|\bs{\varepsilon}\|^2 + \sum_{\ell=1}^L \big[\big(|\hat{\nu}_\ell^{t+1}|^2 +\tau_{\nu_\ell}^{t+1}\big)\big(N- \|\hat{\mb{a}}_\ell^{t+1}\odot\hat{\mb{c}}_\ell^{t+1} \|^2\big) + \tau_{\nu_\ell}^{t+1}\|\hat{\mb{a}}_\ell^{t+1}\odot\hat{\mb{c}}_\ell^{t+1} \|^2\big]\big)$
        \ENDFOR
	\end{algorithmic} 
    \label{algo-2}
\end{algorithm}

% \section{Planar Array response}
% The steering vector in the horizontal and vertical directions are
% \begin{align}
%     \mb{a}_x(\phi,\theta) &= \big[1,\e^{\ji kd_x\sin\theta\cos\phi}, \ldots, \e^{\ji k(M-1) d_x \sin\theta \cos\phi} \big]^T \\
%     \mb{a}_y(\phi,\theta) &= \big[1,\e^{\ji kd_y\sin\theta\sin\phi}, \ldots, \e^{\ji k(N-1) d_y\sin\theta \sin\phi} \big]^T.
% \end{align}
% where $\theta$ and $\phi$ are the azimuth and elevation of the DoA, $k = 2\pi/\lambda$ is the wavenumber and $\lambda$ is the carrier wavelength. The steering vector of an $M\times N$ array is represented by
% $$\mb{a}(\phi,\theta) = \mb{v}_x(\phi,\theta) \otimes \mb{v}_y(\phi,\theta).$$

\vspace{-0.2cm}

\subsection{Proposed SG-VB for UPA Channel Estimation}
Similar to ULA case, let $\gamma\triangleq 1/N_0$ denote the precision floated as an unknown random variable that needs
to be estimated by the VB framework.
The main goal is to infer the distributions of DoAs, $\bs{\theta} = [\theta_1,\ldots,\theta_L]^T$, $\bs{\varphi} = [\varphi_1,\ldots,\varphi_L]^T$ and distance, $\mb{r} = [r_1,\ldots,r_L]^T$, through the variables $\bs{\omega} = [\omega_1,\ldots,\omega_L]^T$, $\bs{\psi} = [\psi_1,\ldots,\psi_L]^T$ and $\mb{s}=[s_1,\ldots,s_L]^T$, respectively, 
given the observation $\mb{y}$. To accomplish this, we employ the mean-field variational distribution $q(\bs{\alpha},\bs{\omega},\bs{\psi},\mb{s},\gamma)$, such that
\begin{align} \label{eq:mean-field}
p(\bs{\alpha}, \bs{\omega}, \bs{\psi}, \mb{s}, &\bs{\beta}, \gamma \mid \mb{y})\nonumber\\
&\approx q(\bs{\alpha}, \bs{\omega}, \bs{\psi}, \mb{s}, \bs{\beta}, \gamma) \nonumber \\
&= q(\bs{\alpha}) \, q(\bs{\omega}) \, q(\bs{\psi}) \, q(\mb{s}) \, q(\bs{\beta}) \, q(\gamma) \nonumber \\
&= q(\gamma) \prod_{\ell=1}^L q(\alpha_{\ell}) \, q(\omega_{\ell}) \, q(\psi_{\ell}) \, q(s_{\ell}) \, q(\beta_{\ell}).
\end{align}
The joint distribution $p(\mb{y},\bs{\nu},\bs{\omega},\mb{s},\bs{\beta},\gamma)$ of the observed variables $\mb{y}$ and the latent variables $\bs{\alpha},\bs{\omega},\bs{\psi},\mb{s},\bs{\beta},\gamma$ can be factorized as
\begin{align} \label{eq:jointPDF}
p(\mb{y},\bs{\alpha},\bs{\omega},\bs{\psi},\mb{s},\bs{\beta},\gamma) = p(\mb{y}|\bs{\alpha},\bs{\omega},\bs{\psi},\mb{s},\gamma) p(\bs{\alpha}|\bs{\beta})p(\bs{\omega}) \nonumber \\
\times p(\bs{\psi})p(\mb{s})p(\bs{\beta})p(\gamma).
\end{align}

Here, $\alpha_{\ell}$ is assumed to be drawn from a zero-mean Gaussian distribution, i.e., $p(\alpha_{\ell}|\beta_{\ell}) = \mc{CN}\big(\alpha_{\ell};0,\beta_{\ell}^{-1}\big)$, where the variance $\beta_{\ell}$ follows Gamma distribution $p(\beta_{\ell}) = \Gamma(a_{\beta},b_{\beta})$.

\emph{2a.) Updating $\omega_{\ell}$:} 
Taking the expectation of the conditional
\eqref{eq:jointPDF} w.r.t. all latent variables except $\omega_{\ell}$, the variational distribution $q(\omega_{\ell})$ can be obtained as
\begin{align}
    q(\omega_\ell) &\propto p(\omega_\ell) \exp\big\{\lr{\ln p(\mb{y}|\bs{\alpha},\bs{\omega},\bs{\psi},\mb{s},\gamma)}_{-\omega_\ell}\big\}\nonumber\\
    &\propto p(\omega_\ell) \exp\Bigg\{-\bblr{\gamma\bigg\|\mb{y} - \sum_{\ell=1}^L \alpha_\ell \mb{a}(\omega_\ell) \odot \mb{c}(\psi_\ell) \nonumber\\
    &\qquad\qquad\qquad\qquad\qquad\qquad\qquad\qquad\odot \mb{d}(s_\ell)\bigg\|^2}_{-\omega_\ell}\Bigg\} \nonumber \\
    &\propto p(\omega_\ell) \exp\Bigg\{-\bblr{\gamma\bigg\|\mb{y} - \sum_{i\neq \ell}^L \alpha_i \mb{a}(\omega_i) \odot \mb{c}(\psi_i) \odot \mb{d}(s_i) \nonumber\\
    &\qquad\qquad\qquad - \bigg(\alpha_\ell \mb{a}(\omega_\ell) \odot \mb{c}(\psi_\ell) \odot \mb{d}(s_\ell)\bigg)\bigg\|^2}_{-\omega_\ell}\Bigg\} \nonumber \\
    &\propto p(\omega_\ell) \exp\big\{\Re\big\{\bs{\eta}_\ell^H \mb{a}(\omega_\ell)\big\} \big\},
\end{align}
where 
\begin{align} \label{eq:eta_i}
\bm{\eta}_{\ell}
&= 2\hat{\gamma}_\ell \hat{\alpha}^*_{\ell} \odot 
    \hat{\mathbf{c}}^*(\psi_{\ell}) \odot 
    \hat{\mathbf{d}}^*(s_{\ell}) \nonumber \\
&\quad\odot 
    \bigg(
        \mathbf{y} 
        - \sum_{i \neq \ell}^{L} 
            \alpha_{i} \, 
            \hat{\mathbf{a}}(\omega_{i}) \odot 
            \hat{\mathbf{c}}(\psi_{i}) \odot 
            \hat{\mathbf{d}}(s_{i})
    \bigg)
\end{align}

 \emph{2b.) Updating $\psi_{\ell}$:} 
Taking the expectation of the conditional
\eqref{eq:jointPDF} w.r.t. all latent variables except $\psi_{\ell}$, the variational distribution $q(\omega_{k\ell})$ can be obtained as
\begin{align}
    q(\psi_\ell) &\propto p(\psi_\ell) \exp\big\{\lr{\ln p(\mb{y}|\bs{\alpha},\bs{\omega},\bs{\psi},\mb{s},\gamma)}_{-\omega_\ell}\big\}\nonumber\\
    &\propto p(\psi_\ell) \exp\Bigg\{-\bblr{\gamma\bigg\|\mb{y} - \sum_{\ell=1}^L \alpha_\ell \mb{a}(\omega_\ell) \odot \mb{c}(\psi_\ell) \nonumber\\
    &\qquad\qquad\qquad\qquad\qquad\qquad\qquad\odot \mb{d}(s_\ell)\bigg\|^2}_{-\psi_\ell}\Bigg\} \nonumber \\
    &\propto p(\psi_\ell) \exp\Bigg\{-\bblr{\gamma\bigg\|\mb{y} - \sum_{i\neq \ell}^L \alpha_i \mb{a}(\omega_i) \odot \mb{c}(\psi_i) \odot \mb{d}(s_i) \nonumber\\
    &\qquad\qquad\qquad - \alpha_\ell \mb{a}(\omega_\ell) \odot \mb{c}(\psi_\ell) \odot \mb{d}(s_\ell)\bigg\|^2}_{-\psi_\ell}\Bigg\} \nonumber \\
    &\propto p(\omega_\ell) \exp\big\{\Re\big\{\bs{\lambda}_\ell^H \mb{c}(\psi_\ell)\big\} \big\},
\end{align}
where 
\begin{align} \label{eq:eta_i}
     \bm{\lambda}_{\ell}
     &= 2\hat{\gamma}_\ell\hat{\alpha}^*_{\ell} \odot
    \hat{\mathbf{a}}^*(\omega_{\ell}) \odot 
    \hat{\mathbf{d}}^*(s_{\ell}) \nonumber \\
    &\quad\odot
    \bigg(\mb{y} - \sum_{i\neq \ell}^{L} \alpha_{i} \, 
    \hat{\mathbf{a}}(\omega_{i}) \odot 
    \hat{\mathbf{c}}(\psi_{i}) \odot 
    \hat{\mathbf{d}}(s_{i})\bigg)
 \end{align}

\emph{2c.) Updating $\s_{\ell}$:} 
Taking the expectation of the conditional
\eqref{eq:jointPDF} w.r.t. all latent variables except $\s_{\ell}$, the variational distribution $q(\s_{\ell})$ can be obtained as
\begin{align}
    q(\s_\ell) &\propto p(\s_\ell) \exp\big\{\lr{\ln p(\mb{y}|\bs{\alpha},\bs{\omega},\bs{\psi},\mb{s},\gamma)}_{-\s_\ell}\big\}\nonumber\\
    &\propto p(\s_\ell) \exp\Bigg\{-\bblr{\gamma\bigg\|\mb{y} - \sum_{\ell=1}^L \alpha_\ell \mb{a}(\omega_\ell) \odot \mb{c}(\psi_\ell) \nonumber \\
    &\quad\quad\quad\quad\quad\quad\quad\quad\quad\quad\quad\quad\quad\quad\quad\odot \mb{d}(s_\ell)\bigg\|^2}_{-\s_\ell}\Bigg\} \nonumber \\
    &\propto p(\s_\ell) \exp\Bigg\{-\bblr{\gamma\bigg\|\mb{y} - \sum_{i\neq \ell}^L \alpha_i \mb{a}(\omega_i) \odot \mb{c}(\psi_i) \odot \mb{d}(s_i) \nonumber\\
    &\qquad\qquad\qquad - \alpha_\ell \mb{a}(\omega_\ell) \odot \mb{c}(\psi_\ell) \odot \mb{d}(s_\ell)\bigg\|^2}_{-\psi_k}\Bigg\} \nonumber \\
    &\propto p(s_\ell) \exp\big\{\Re\big\{\bs{\zeta}_\ell^H \mb{d}(s_\ell)\big\} \big\},
\end{align}
where 
\begin{align} \label{eq:eta_i}
     \bm{\zeta}_{\ell}
     &= 2\hat{\gamma}_\ell\hat{\alpha}^*_{\ell} \odot
    \hat{\mathbf{a}}^*(\omega_{\ell}) \odot 
    \hat{\mathbf{c}}^*(\psi_{\ell})\nonumber \\
    &\quad\odot\bigg(\mb{y} - \sum_{i\neq \ell}^{L} \alpha_{i} \, 
    \hat{\mathbf{a}}(\omega_{i}) \odot 
    \hat{\mathbf{c}}(\psi_{i}) \odot 
    \hat{\mathbf{d}}(s_{i})\bigg)
 \end{align}

 Similar to the ULA case, to find the optimal $s_\ell$, we maximize the log-likelihood of $s_\ell$ given the residual correlation:
  \begin{align}
    [\mc{L}_\ell]_k = \Re\{{\bs{\zeta}}_\ell^H \mb{d}(s_k)\},
\end{align}
where $s_k \in \mc{S}_{\rm grid}$ for $k = 1,\ldots,K$, and $K$ denotes the number of coarse grid points in the set $\mc{S}_{\rm grid}$. The estimate of $s_\ell$ iteratively using the Newton-Raphson update:
\begin{align}
    s_\ell^{t+1} = s_\ell^t - \epsilon\frac{\Delta\mc{L}(s_\ell^t)}{\Delta^2\mc{L}(s_\ell^t)},
\end{align}
where the step size $\epsilon$ is set to $0.01$ and the first and second derivatives of $[\mc{L}_\ell]_k$ w.r.t. $s_\ell$ can be derived, respectively, as
\begin{align}
    \Delta[\mc{L}_\ell]_k &= -\Im\{{\bs{\zeta}}_\ell^H (\mb{m}\odot\mb{c}(s_k))\},\\
    \Delta^2 [\mc{L}_\ell]_k &= \Re\{{\bs{\zeta}}_\ell^H (\mb{m}^2\odot\mb{c}(s_k))\},
\end{align}
where the index vector is defined as $\mb{m} \triangleq [e^{-j s \left(m_1^2+n_1^2 \right)},\ldots,e^{-j s \left(m_M^2+n_M^2 \right)}]^T \in \mathbb{C}^N$.

\emph{2d.) Update ${\alpha}_{\ell}$:} 
Taking the expectation of the conditional \eqref{eq:jointPDF} w.r.t. all latent variables except ${\alpha}_{\ell}$, the variational distribution $q(\alpha_{\ell})$ can be obtained as
\begingroup
\allowdisplaybreaks
\begin{align}
    q(\alpha_\ell) &\propto \exp\big\{\lr{\ln p(\mb{y}|\bs{\alpha},\bs{\omega},\bs{\psi},\mb{s},\gamma) + \ln p(\alpha_\ell|\beta_\ell)}_{-\alpha_\ell}\big\} \nonumber \\
    &\propto \mc{CN}\bigg\{-\hat{\gamma}\big(\big\|\hat{\mb{a}}_\ell\odot\hat{\mb{c}}_\ell\odot\hat{\mb{d}}_\ell\big\|^2|\alpha_\ell|^2 \nonumber \\
    &\quad - 2\,\Re\bigg\{\alpha_\ell \bigg(\mb{y} -\sum_{i\neq \ell}^L\hat{\alpha}_i \hat{\mb{a}}_i \odot \hat{\mb{c}}_i \odot \hat{\mb{d}}_i \bigg)^H \nonumber\\
    &\qquad\qquad\times\big(\hat{\mb{a}}_\ell\odot\hat{\mb{c}}_\ell \odot\hat{\mb{d}}_\ell\big)\bigg\} - \lr{\beta_\ell} |\alpha_\ell|^2\bigg\} \nonumber \\
    &\propto \mc{CN}\big(\alpha_\ell; z_\ell, (M\hat{\gamma})^{-1}\big) \mc{CN}\big(\alpha_\ell;0,\hat{\beta}_\ell^{-1}\big) \nonumber \\
    &\propto \mc{CN}\big(\alpha_\ell; \hat{\alpha}_\ell, \tau_{\alpha_\ell}\big),
\end{align}
\endgroup
where
%\begin{align}
    $z_\ell \triangleq {(\hat{\alpha}_\ell\big\|\hat{\mb{a}}_\ell\odot\hat{\mb{c}}_\ell\odot\hat{\mb{d}}_\ell\big\|^2 +  \big(\hat{\mb{a}}_\ell\odot\hat{\mb{c}}_\ell\odot\hat{\mb{d}}_\ell\big)^H\bs{\varepsilon})}/N$, \\
%\begin{align}
    ${\tau_{\alpha_\ell} = {(N\hat{\gamma} + \hat{\beta}_\ell)^{-1}}}$, and $
    \hat{\alpha}_\ell = {z_\ell N\hat{\gamma}}{(N\hat{\gamma} + \hat{\beta}_\ell)^{-1}}.$
%\end{align}

\emph{2e.) Update $\beta_{\ell}$:} The variational distribution $q(\beta_{\ell})$ is obtained as 
\begin{align}
    q(\beta_\ell) &\propto p(\beta_\ell) \exp\{\lr{p(\alpha_\ell|\beta_\ell)}\},
\end{align}
Thus, $q(\beta_\ell)$ is a Gamma distribution with mean ${\hat{\beta}_\ell = (|\hat{\alpha_\ell}|^2 + \tau_{\alpha_\ell})^{-1}}$.

\emph{2f.) Update $\gamma_\ell$:} The variational distribution $q(\gamma_\ell)$ can be obtained as 
\begin{align*}
    q(\gamma) &\propto p(\gamma) \exp\big\{\lr{\ln p(\mb{y}|\bs{\alpha},\bs{\omega},\bs{\psi},\mb{s},\gamma)}_{-\gamma}\big\}\\
    &\propto p(\gamma) \exp\Bigg\{\ln\gamma-\gamma\bblr{\bigg\|\mb{y} \!-\! \sum_{\ell=1}^L \alpha_k \mb{a}(\omega_\ell) \!\odot\!\mb{c}(\psi_\ell) \nonumber \\ &\quad\quad\quad\quad\quad\quad\quad\quad\quad\quad\quad\quad\quad\quad\quad\quad\!\odot\! \mb{d}(s_\ell)\bigg\|^2}\Bigg\}.
\end{align*}
Thus, $q(\gamma)$ is a Gamma distribution with mean 
\begin{align}
 \hat{\gamma} = \frac{a_{\gamma} + N}{b_\gamma + \big\|\mb{y} - \sum_{\ell=1}^L \alpha_\ell \mb{a}(\omega_\ell) \odot \mb{c}(\psi_\ell) \odot \mb{d}(s_\ell)\big\|^2}, 
\end{align}
where
\begin{align}
    \bigg\|\mb{y} - \sum_{\ell=1}^L \alpha_\ell \mb{a}&(\omega_\ell) \odot \mb{c}(\psi_\ell) \odot \mb{d}(s_\ell)\bigg\|^2 = \|\bs{\varepsilon}\|^2 + \sum_{\ell=1}^L \tau_{\alpha_\ell}N \nonumber\\
    & + \sum_{\ell=1}^L \big(|\hat{\alpha}_\ell|^2 +\tau_{\alpha_\ell}\big)\big(N - \|\hat{\mb{a}}_\ell\odot\hat{\mb{c}}_\ell\odot\hat{\mb{d}}_\ell \|^2\big).
\end{align}

By iteratively updating $\omega_\ell, \psi_\ell, s_\ell, \alpha_\ell,$ and $\gamma$, we derive the SG-VB algorithm for the UPA near-field channel estimation problem. The implementation of SG-VB for UPA closely follows that for ULA; hence, we omit a separate summary of the UPA implementation.

\subsection{Computational Complexity Analysis}
We conclude this section by presenting the computational complexities of the proposed SG-VB, MUSIC\cite{stoica2005spectral}, P-SOMP, P-SIGW \cite{cui2022channel}, and SBL\cite{fang2014pattern} methods, as summarized in Table \ref{tab:complexity}.
\begin{table*}[th]
\centering
\begin{tabular}{|c|c|c|} 
 \hline
 \bf{Algorithm} & \bf{Complexity} \\ 
 \hline\hline
The proposed SG-VB & $\mathcal{O}(L(N^2 + KN)$  \\ \hline
MUSIC \cite{stoica2005spectral} & $\mathcal{O}(N^3 + K_aK_r)$  \\ \hline
P-SOMP, P-SIGW \cite{cui2022channel} & $\mc{O}(LN^2S)$ \\ \hline
SBL \cite{fang2014pattern} & $\mc{O}(N^3)$  \\ \hline
\end{tabular}
\caption{Algorithm Computational Complexity}
\label{tab:complexity}
\end{table*}
The complexity of MUSIC is $\mathcal{O}(N^3 + K_aK_r)$, where $K_a$ and $K_r$ denote the numbers of angle and distance samples, respectively. 
The computational complexities of P-SOMP and P-SIGW are $\mathcal{O}(LN^2S)$, where $S$ is the number of distance rings. 
The computational complexity of SBL is $\mathcal{O}(N^3)$. 
For the proposed method SG-VB, the computational complexities of Steps 6, 7, 8, and 11 are $\mathcal{O}(N)$, $\mathcal{O}(N^2)$, $\mathcal{O}(N^2)$, and $\mathcal{O}(NK)$, respectively. 
Thus, the total computational complexity of Algorithm~2 is $\mathcal{O}\!\big(L(N^2 + KN)\big)$. Although the proposed SG-VB method has higher computational complexity than P-SOMP, it achieves significantly better channel estimation performance.

\section{Simulation Results}
This section presents numerical results demonstrating the performance of the proposed SG-VB method for near-field channel estimation with both ULA and UPA.

For performance comparison, we consider the following benchmarks:
\begin{itemize}
    \item The least-squared (LS) method.
    \item Oracle LS, which assumes exact knowledge of the scatterers’ or users’ angles and distances, and thus serves as a performance bound.
    \item P-SOMP \cite{cui2022channel}, which exploits the sparsity of the channel in polar-domain.
    \item SBL \cite{fang2014pattern}, which utilizes the same codebook as P-SOMP to transform the channel into its polar-domain representation, followed by Sparse Bayesian Learning to recover the sparse channel vector.
\end{itemize}

\subsection{ULA Channel Estimation}
The simulation parameters for the ULA are summarized in Table \ref{tab:SimPara_ula}. Note that the complex gain $\alpha_\ell$ are assumed to follow a Gaussian distribution with zero mean and unit variance. In addition to LS and Oracle LS, we compare the performance of the proposed SG-VB with existing methods such as MUSIC \cite{he2011efficient}, P-SIGW \cite{cui2022channel}. Beyond channel estimation, we also evaluate angle and distance recovery performance, defined as:

\begin{itemize}
    \item The mean squared error (MSE) of angles estimation, defined as  
    \[
    {\rm MSE}(\hat{\bs{\theta}}) \!=\! 10\log\big(\| \bs{\theta}-\hat{\bs{\theta}}\|^2\big),
    \]  
    where $\hat{\bs{\theta}} = \arcsin(\hat{\bs{\omega}}/\pi)$ denotes the estimated angles.  

    \item The NMSE of distance estimation, defined as  
    \[
    {\rm NMSE}(\hat{\mb{r}}) \!=\! 10\log\left(\frac{\| \mb{r}-\hat{\mb{r}}\|^2}{\|\mb{r}\|^2}\right),
    \]  
    where $\hat{\mb{r}} = \pi\lambda/(4\hat{\mb{s}})$ denotes the estimated distance vector.  
\end{itemize}

\begin{table}[!t]
			\caption{Simulation Configurations for ULA}
			\label{tab:SimPara_ula}
			\centering
			\begin{tabular}{|p{17em}|p{8em}|}
				\hline
				The number of array antennas $N$         &  256     \\ \hline                
				Carrier frequency    $f_c$ & 100 GHz \\ \hline
				Minimum allowable distance $r_{\text{min}}$ & 3 meters \\ \hline
				Maximum allowable distance $r_{\text{max}}$ & 90 meters \\ \hline
                The number of channel paths $L$ & 6 \\ \hline 
				The distribution of $\theta$ &  $\mathcal{U}\big(-60^\circ , 60^\circ\big)$ \\ \hline 
				Signal-to-noise ratio SNR & $L / N_0$ \\ \hline
				% Dictionary size of DoA & $2N$ \\ \hline 
    %             Dictionary size of angle-distance &  12$N$ \\ \hline 
				Maximum number of iterations of SG-VB & 150 \\ \hline
                Angle-distance sampling (dictionary size) for MUSIC, P-SOMP and P-SIGW & $3N$ \\ \hline
                % Angle-distance resolutions of P-SOMP and P-SIGW & $2N$ and $4.3N$ \\ \hline
			\end{tabular}
		%\vspace*{-2em}
		\end{table}

\begin{figure}[!th]
	\centering
\includegraphics[width=1.0\linewidth]{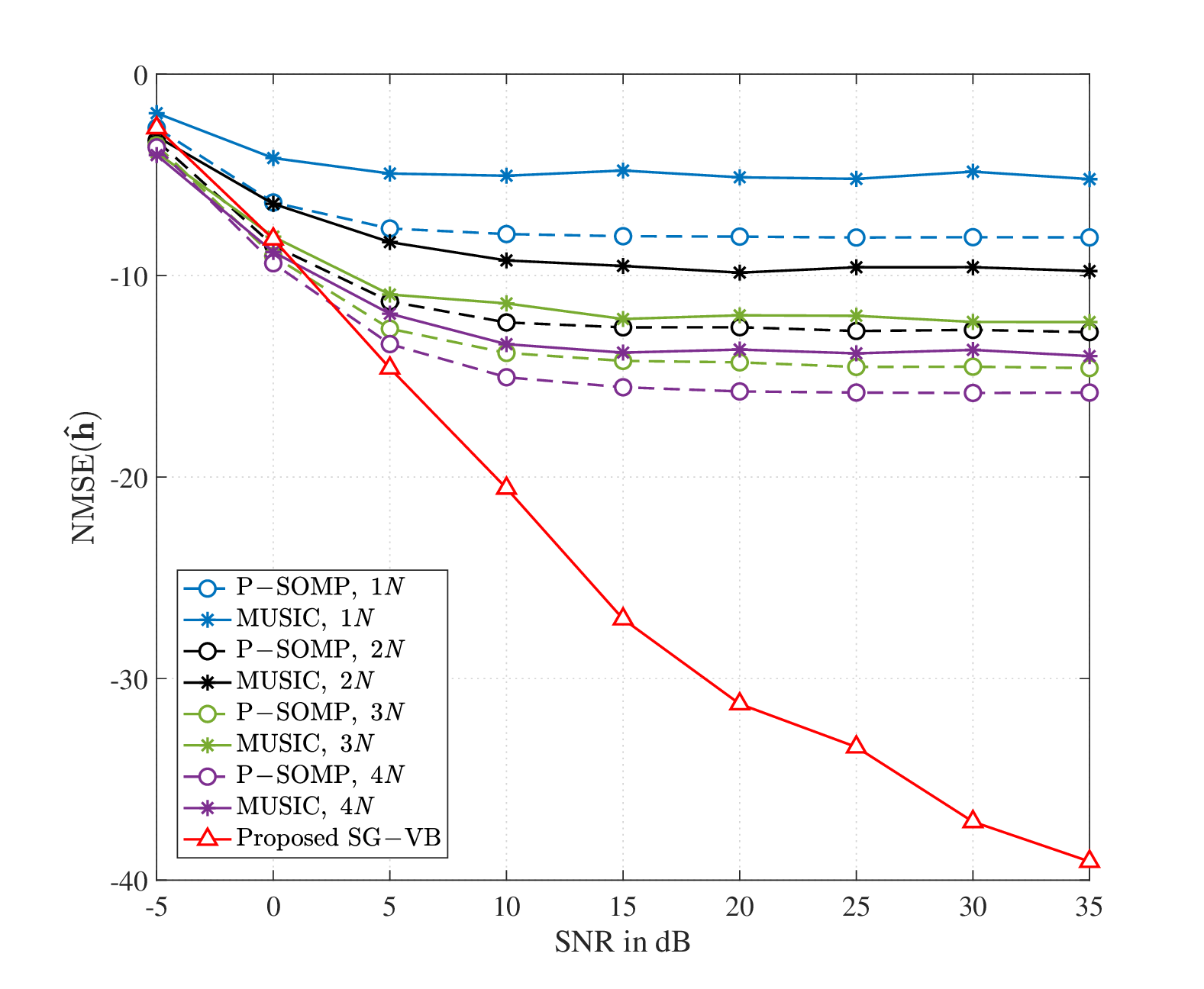}
	\caption{NMSE vs. SNR for different grid size.}
	\label{ula_NMSE_SNR_grid_size}
\end{figure}

Fig. \ref{ula_NMSE_SNR_grid_size} shows the NMSE versus SNR for different grid sizes ranging from $1N$ to $4N$. It can be observed that the performance improves as the grid size increases. However, the performance gain becomes limited when the grid size increases from $3N$ to $4N$. This can be attributed to the fact that an excessively dense grid leads to high correlation among dictionary atoms, resulting in limited performance improvement \cite{codebooksize}. We therefore choose the dictionary size to be $3N$ in order to balance performance and complexity.

\begin{figure}[!th]
	\centering
\includegraphics[width=1.0\linewidth]{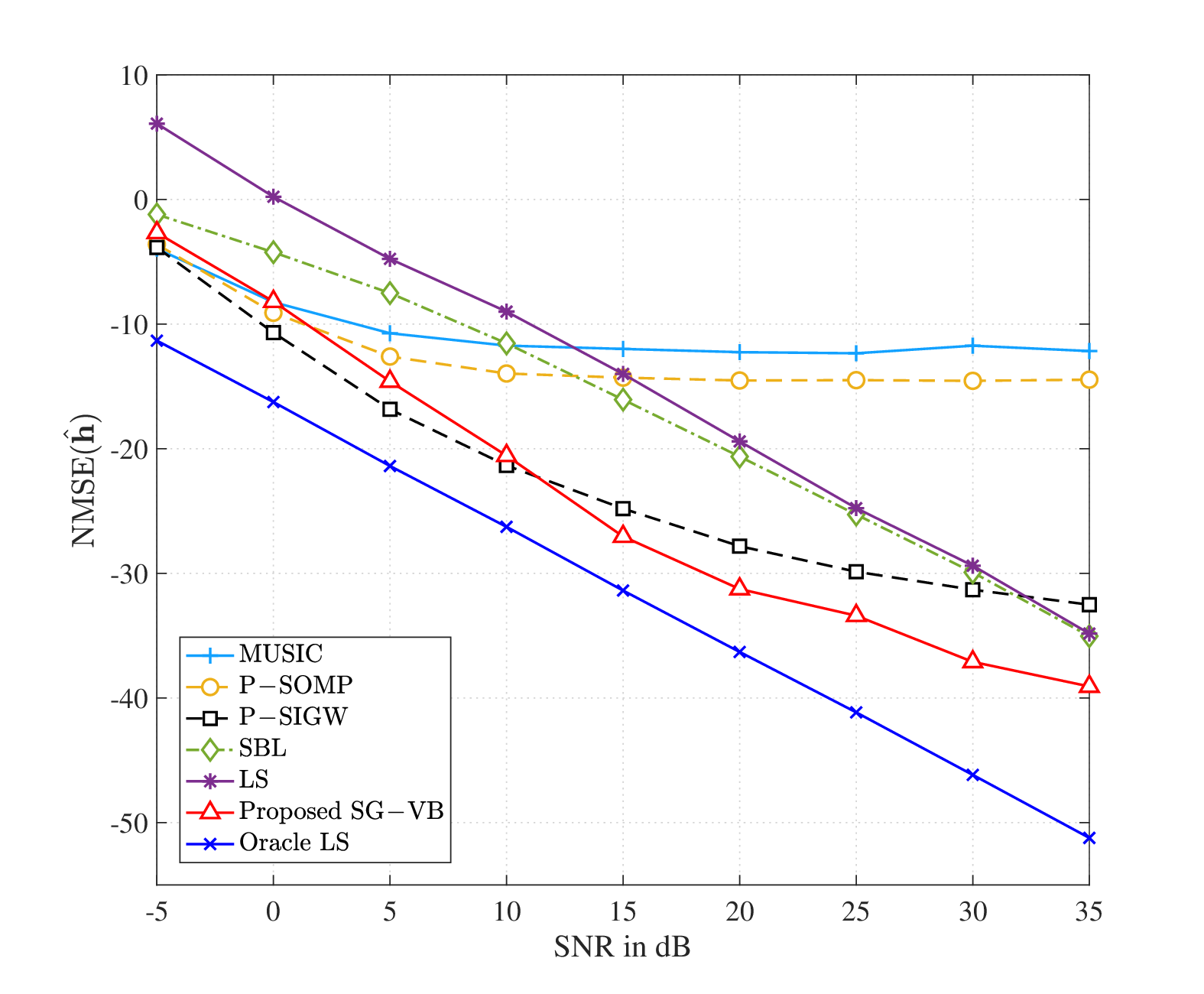}
	\caption{NMSE vs. SNR when $L=6$ and $N=256$.}
	\label{ula_NMSE_SNR}
\end{figure}

Fig. \ref{ula_NMSE_SNR} shows the NMSE performance of all algorithms  as a function of the SNR. The proposed SG-VB consistently outperforms MUSIC, P-SOMP, P-SIGW, LS, and SBL in reconstructing the near-field channel, especially at high SNRs. In the low-SNR regime, P-SIGW performs slightly better than the proposed SG-VB; however, this comes at the cost of an excessively large dictionary. The performance saturation observed in MUSIC- or OMP-based methods can be attributed to their reliance on a fixed grid, which limits estimation accuracy. The SBL algorithm achieves performance comparable to LS, but its effectiveness is affected by the group-element sparsity prior, which is not always guaranteed. In contrast, the proposed SG-VB adopts a gridless approach to estimate the continuous DoAs, yielding performance close to the Oracle LS lower bound.

\begin{figure}[!th]
	\centering
\includegraphics[width=1.0\linewidth]{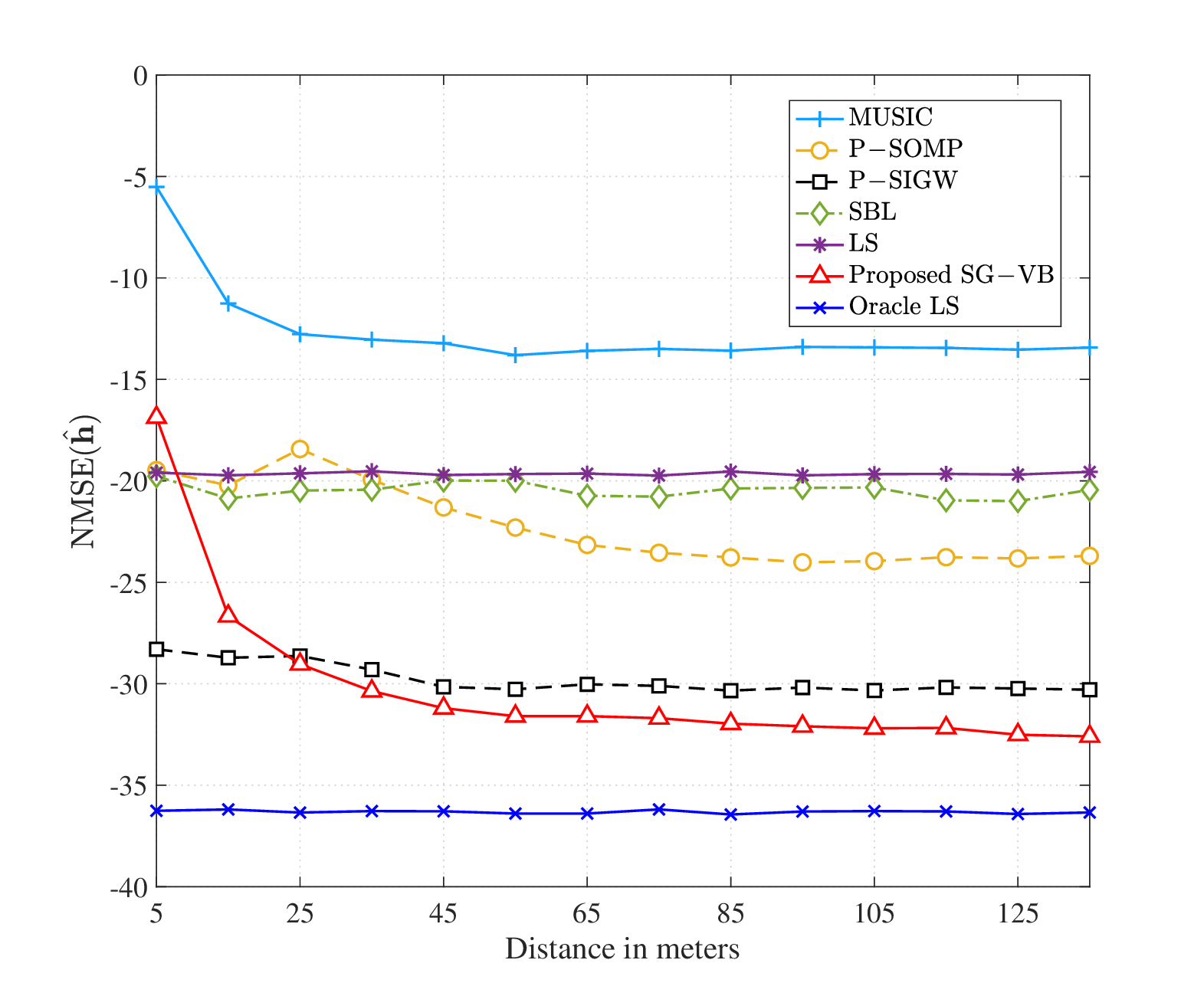}
	\caption{NMSE vs. Distance for ULA with $L=6$, SNR = 20 dB, and $N=256$.}
	\label{ula_NMSE_Distance}
\end{figure}

Fig. \ref{ula_NMSE_Distance} illustrates the NMSE performance of all algorithms with respect to distance between the user and the center of the array, ranging from 5 to 135 meters with step size of 10 meters. The Rayleigh distance is around $98.3$ meters. Each path is independently modeled as a virtual source whose distance to the BS is randomly drawn from $[r_{\min}, r_{\max}]$. In addition, each path has its own randomly generated angle $\theta_\ell$, so the sources are not constrained to lie on the same line; the model does not enforce any explicit relationship to a fixed user--BS distance. Although the proposed SG-VB algorithm has a lower performance in comparison with that of off-grid P-SIGW at distances below $15$ meters, it achieves excellent performance across the distance range by reducing the gap with the lower bound Oracle LS. In addition, the high estimation performance of P-SIGW comes at a high computational, since the codebooks of P-SOMP, P-SIGW, and SBL use an angular resolution of $3N$, with about $4.3$ non-uniformly sampled distances per angle, totaling roughly 3302 angle–distance points.
\begin{figure}[!t]
	\centering
\includegraphics[width=1.0\linewidth]{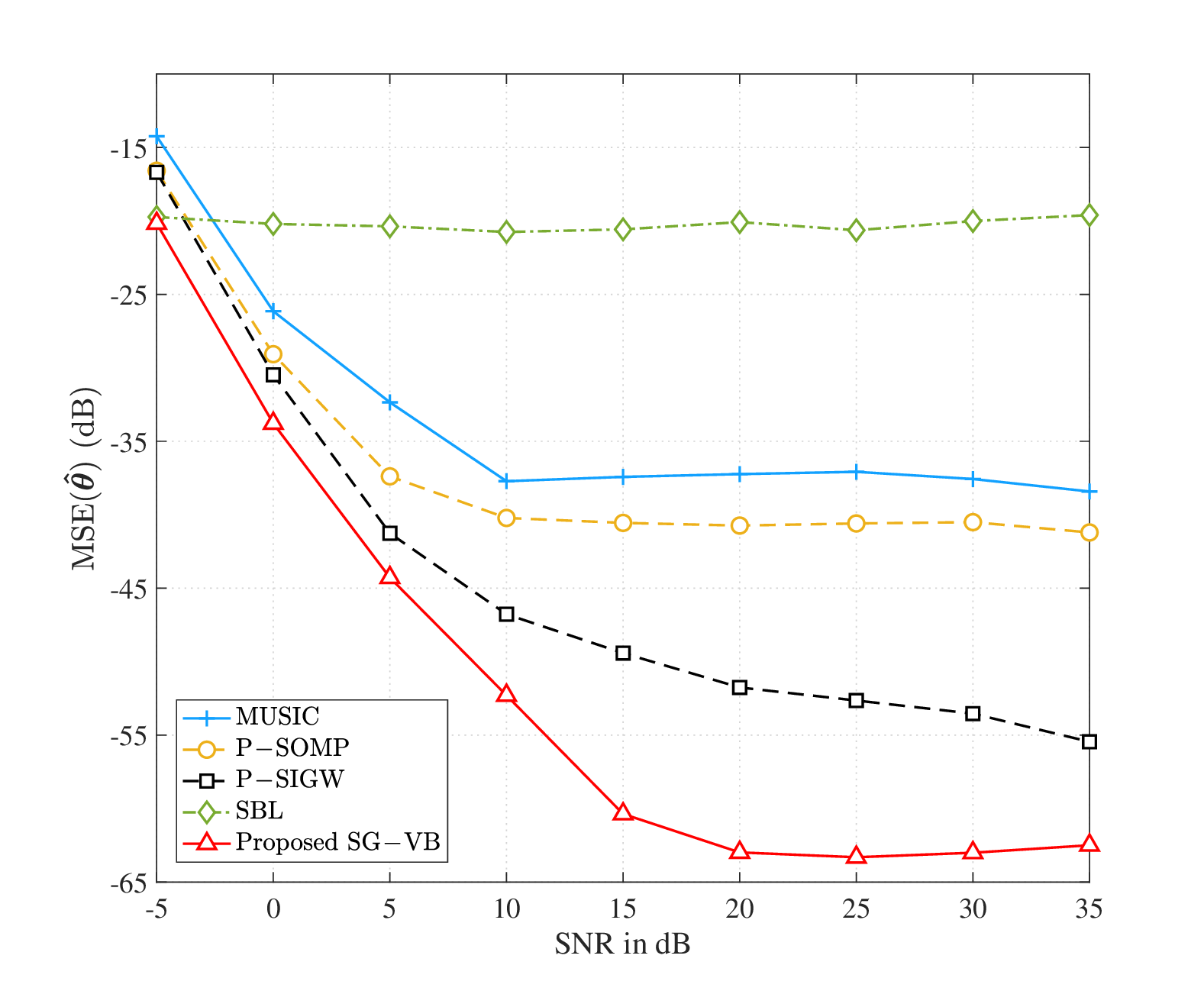}
	\caption{MSE of angle vs. SNR for ULA with $L=6$, $N=256$.} 
	\label{ula_MSE_SNR_Angle}
\end{figure}
%\vspace{-0.4cm}

Fig. \ref{ula_MSE_SNR_Angle} illustrates the angle estimation performance of all algorithms as a function of the SNR. It is clear that the proposed SG-VB outperforms all baselines across almost the entire SNR range, particularly at SNRs above $10$ dB, thanks to its gridless VB approach. At SNRs below $10$ dB, grid-based methods, including P-SOMP, P-SIGW, have slightly better performance than SG-VB; however, they suffer from performance saturation when the true DoAs deviate from the discretized angular grid, which limits their estimation accuracy at high SNR.

\begin{figure}[!t]
	\centering
\includegraphics[width=1.0\linewidth]{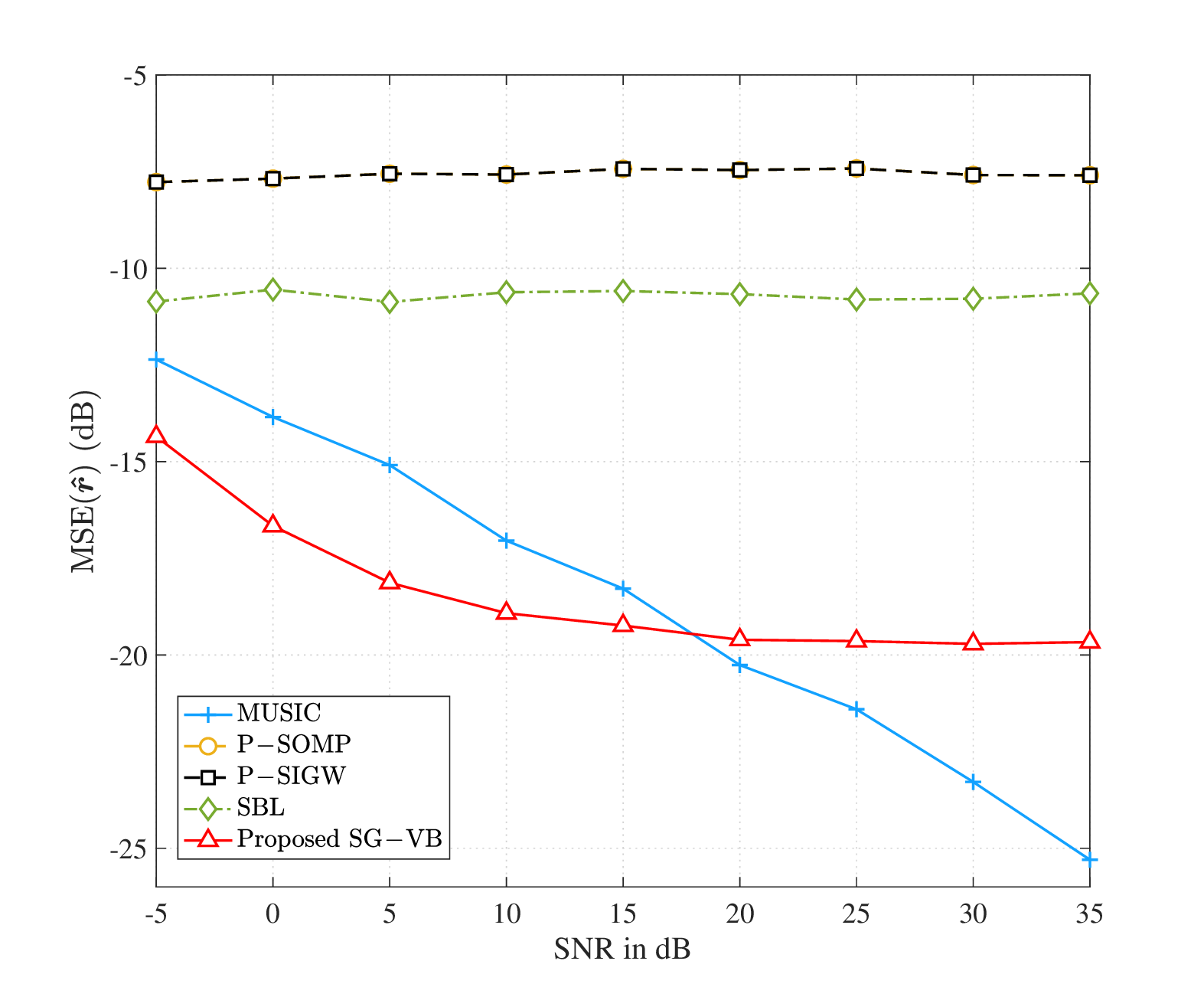}
	\caption{NMSE of distance vs. SNR for ULA with $L=6$, $N=256$.}
	\label{ula_NMSE_SNR_distance}
\end{figure}

Fig. \ref{ula_NMSE_SNR_distance} presents the distance estimation performance of all algorithms as a function of the SNR. The proposed SG-VB achieves the best performance among the baselines for SNRs below $20$ dB, but is outperformed by MUSIC at higher SNRs. Interestingly, P-SIGW offers no improvement over P-SOMP. This is because P-SIGW is initialized with coarse distance estimates, while the over-complete angle–distance dictionary of P-SOMP already minimizes quantization error, leaving little room for off-grid refinement by P-SIGW.

\subsection{UPA Channel Estimation}

For UPA channel estimation, in addition to LS, P-OMP, SBL, and Oracle LS, we compare SG-VB with 3D-MUSIC \cite{kosasih2023parametric}. This method employs a two-stage approach: first, estimating the DoAs with a fixed distance; and second, estimating the distance based on the estimated DoAs. Moreover, the codebooks for P-OMP and SBL are generated in the same manner as in \cite{wu2023multiple}. 
% Beyond channel estimation, we also evaluate angle and distance recovery performance, defined as:

\begin{table}[!t]
			\caption{Simulation Configurations for UPA}
			\label{tab:SimPara}
			\centering
			\begin{tabular}{|p{17em}|p{8em}|}
				\hline
				The number antennas per row      $M_H$         &  16     \\ \hline   
                    The number antennas per column     $M_V$         &  16     \\ \hline
				Carrier frequency    $f_c$ & 3 GHz \\ \hline
				Minimum allowable distance $r_{\text{min}}$ & 5 meters \\ \hline
				Maximum allowable distance $r_{\text{max}}$ & 25 meters \\ \hline
                The number of paths $L$ & 3 \\ \hline 
				The distribution of $\theta$ &  $\mathcal{U}\big(-60^\circ , 60^\circ\big)$ \\ \hline
                The distribution of $\varphi$ &  $\mathcal{U}\big(-80^\circ , 80^\circ\big)$ \\ \hline
				Signal-to-noise ratio SNR & $L / N_0$ \\ \hline
				% Dictionary size of DoA & $2N$ \\ \hline 
    %             Dictionary size of angle-distance &  12$N$ \\ \hline 
				Maximum number of iterations of SG-VB & 200 \\ \hline
                % Angle-distance sampling (dictionary size) for P-SOMP and P-SIGW & $2 \times (M_H \times M_V)\times 4.3$ \\ \hline
                % Angle-distance resolutions of P-SOMP and P-SIGW & $2N$ and $4.3N$ \\ \hline
			\end{tabular}
		%\vspace*{-2em}
		\end{table}
        
\begin{figure}[!th]
	\centering
\includegraphics[width=1.0\linewidth]{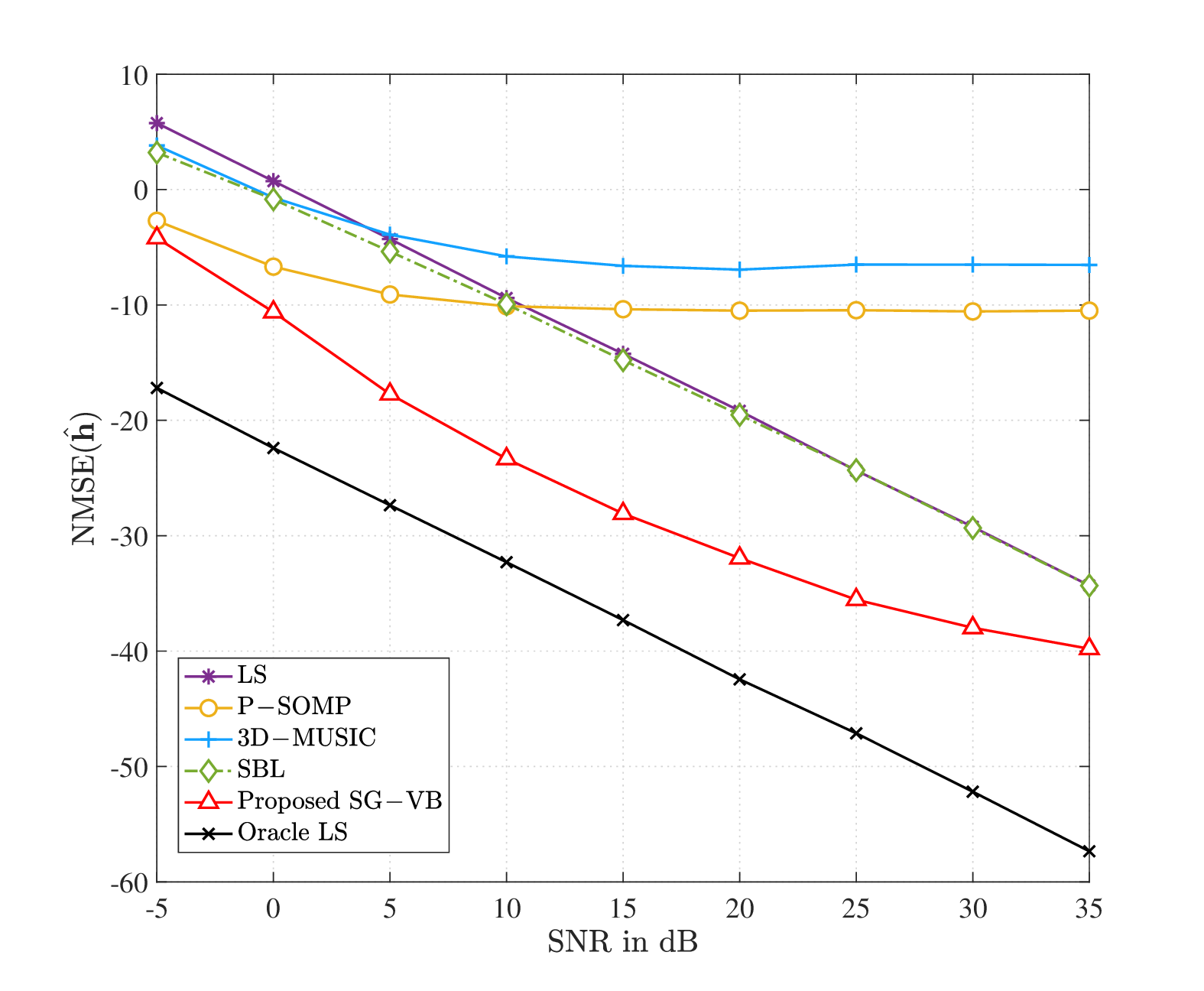}
	\caption{NMSE vs. SNR for UPA with $L=3$, $N=256$.}
	\label{upa_NMSE_SNR}
\end{figure}

Fig. \ref{upa_NMSE_SNR} shows the NMSE performance for near-field channel estimation as a function of the SNR. Similar to the ULA case, the proposed SG-VB outperforms MUSIC, P-SOMP, SBL, and LS across the entire SNR range. These results highlight the effectiveness of SG-VB for near-field channel estimation regardless of array geometry, owing to its gridless DoA estimation and coarse-to-fine distance estimation.

\begin{figure}[!th]
	\centering
\includegraphics[width=1.0\linewidth]{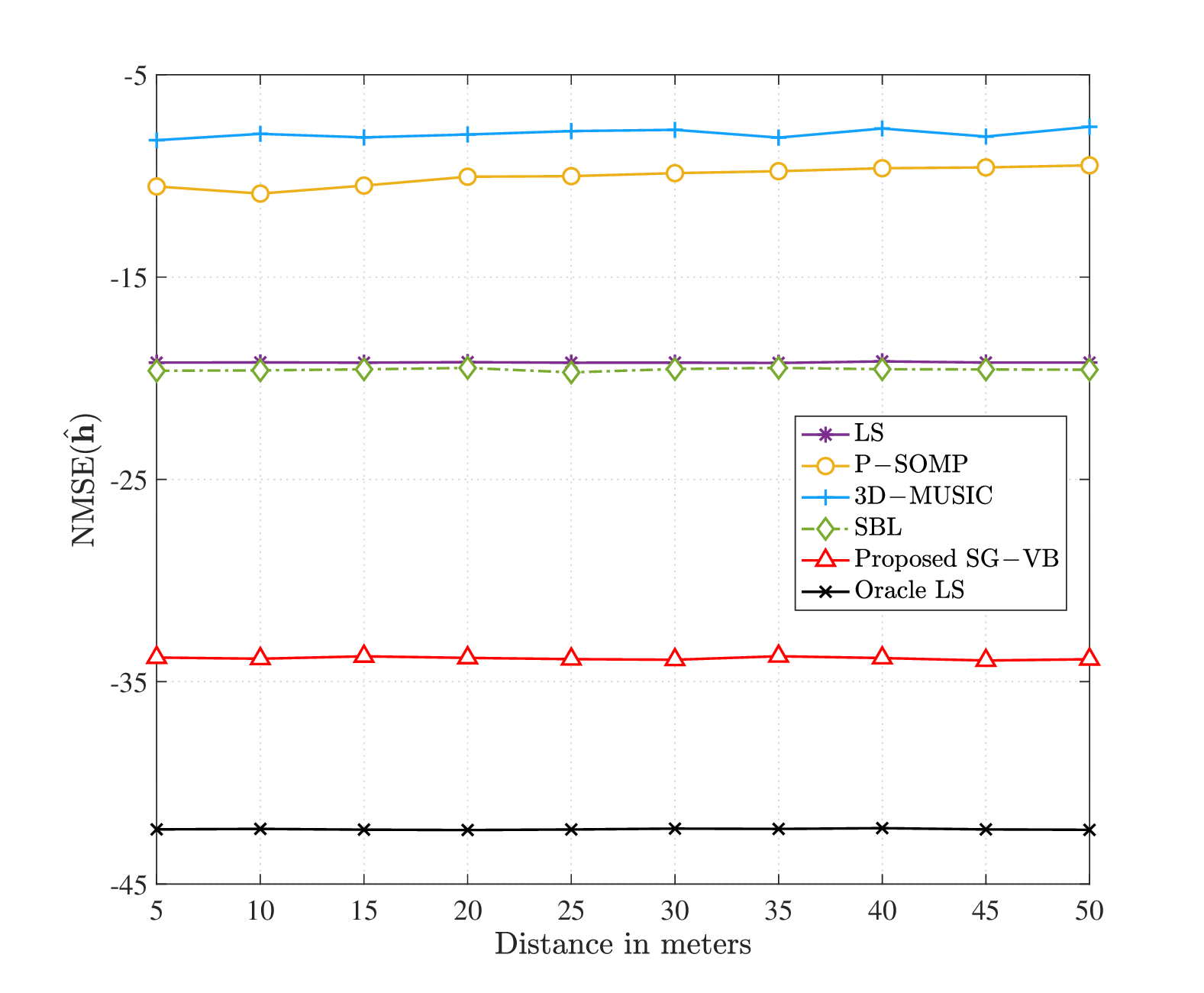}
	\caption{NMSE vs. Distance for UPA with $L=3$, SNR = 20 dB, and $N=256$.}
	\label{upa_NMSE_Distance}
\end{figure}

In Fig. \ref{upa_NMSE_Distance}, we investigate the NMSE near-field channel estimation performance of all algorithms with respect to SNR. As observed, the proposed SG-VB outperforms all methods and achieves the performance that is closest to the performance bound achieved by Oracle LS. In particular, SG-VB provides a constant $14$ dB NMSE gain over LS and SBL, while being far more efficient than 3D-MUSIC and P-SOMP.

\begin{figure}[!t]
	\centering
\includegraphics[width=1.0\linewidth]{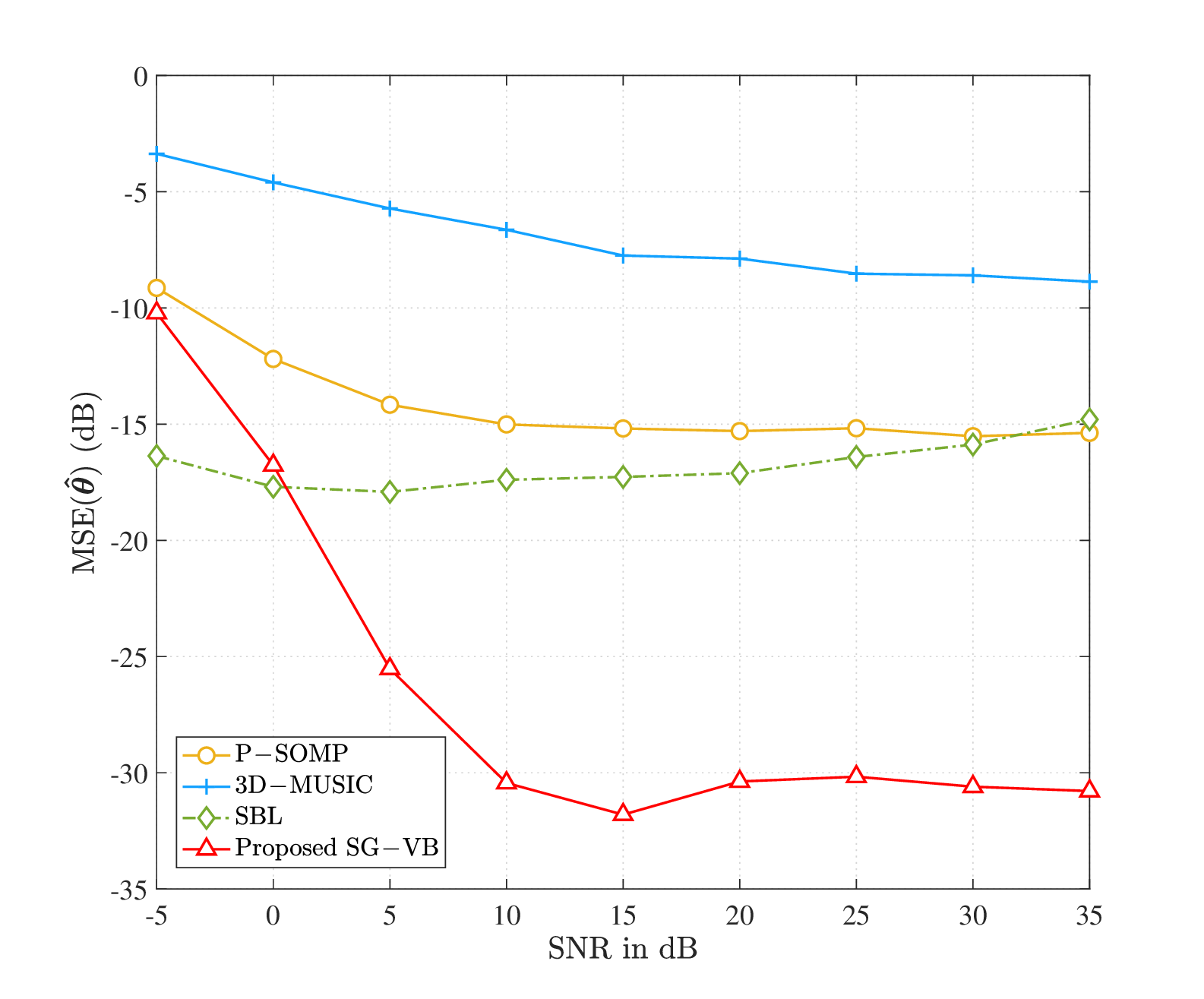}
	\caption{MSE of elevation angle vs. SNR  for UPA with $L=3$, $N=256$.} 
	\label{upa_NMSE_SNR_theta}
\end{figure}

\begin{figure}[!t]
	\centering
\includegraphics[width=1.0\linewidth]{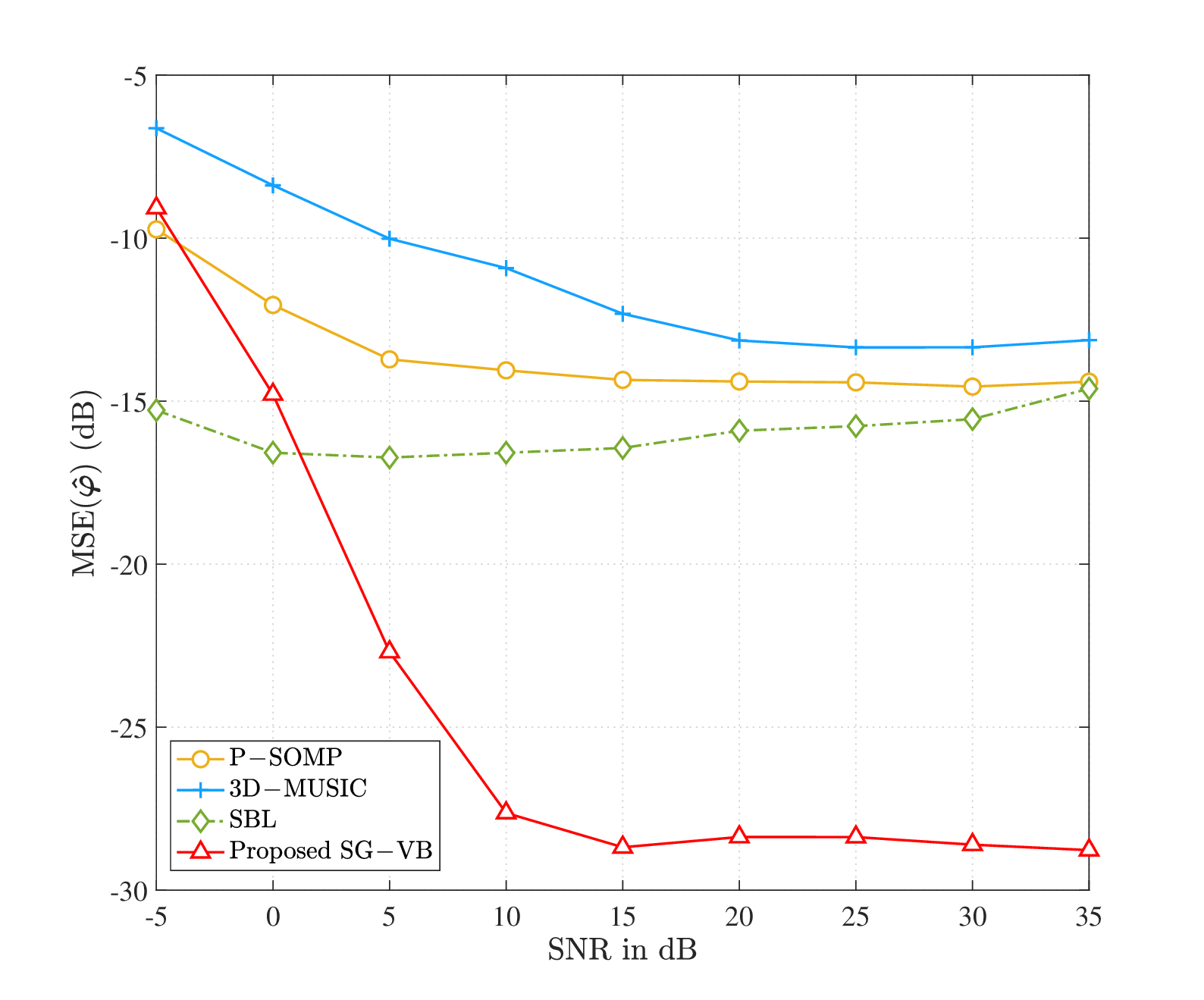}
	\caption{MSE of azimuth angle vs. SNR  for UPA with $L=3$, $N=256$.} 
	\label{upa_MSE_SNR_phi}
\end{figure}

Figs. \ref{upa_NMSE_SNR_theta} and \ref{upa_MSE_SNR_phi} show the angle estimation performance of all algorithms versus SNR. As observed, the proposed SG-VB outperforms the 3D-MUSIC, P-SOMP, and SBL in recovering angles, especially at high SNRs. At low SNRs below $0$ dB, codebook-based methods like P-SOMP and SBL achieve a better performance compared to the proposed SG-VB; however, their performance quickly saturates as SNR increases. This is due to the overcomplete dictionary, leaving little room for improvement at high SNR. In contrast, the gridless angle estimation of SG-VB achieves an approximate $20$ dB gain at high SNRs, demonstrating the superiority of the proposed method.

\begin{figure}[!t]
	\centering
\includegraphics[width=1.0\linewidth]{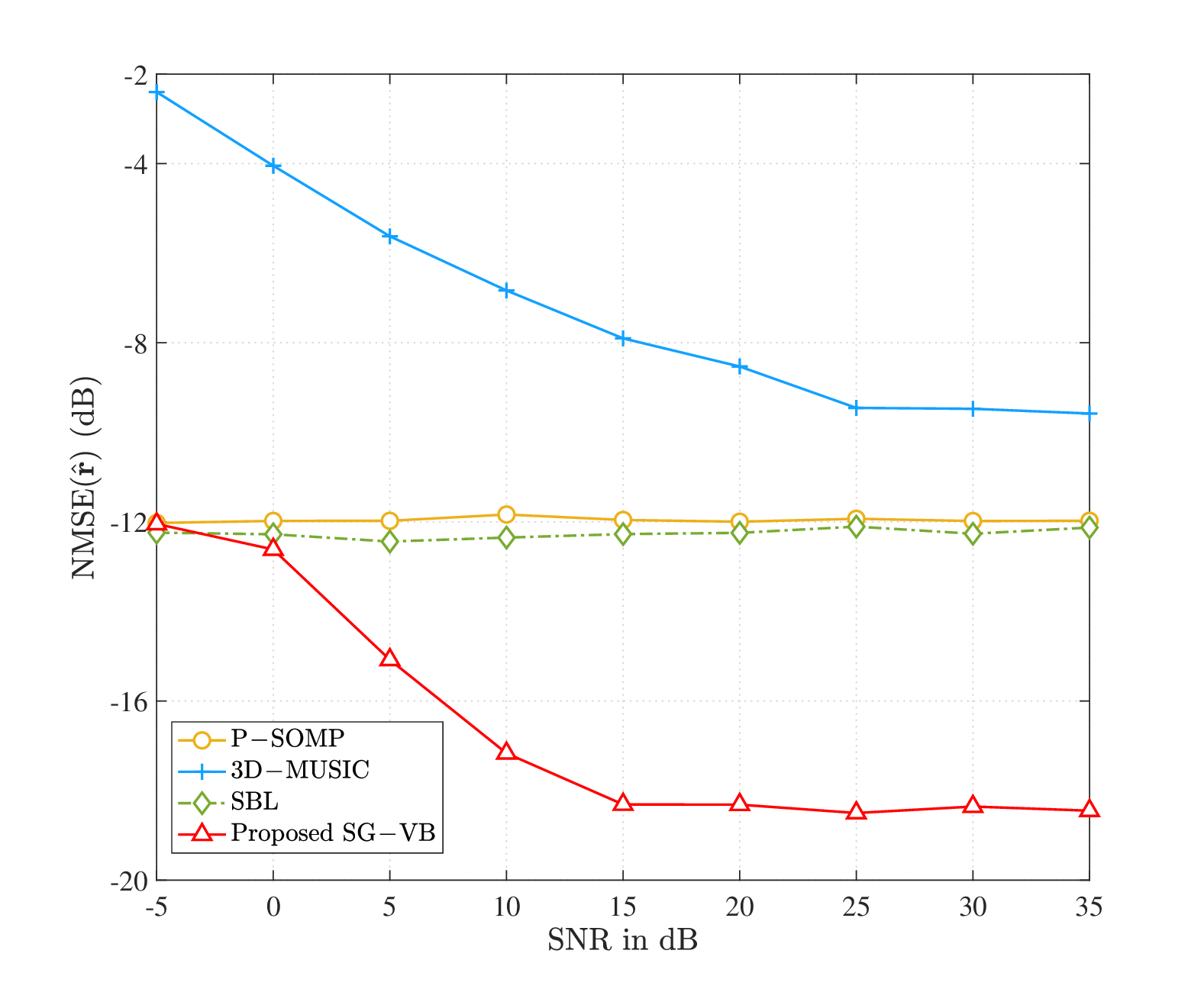}
	\caption{NMSE of distance vs. SNR for UPA with $L=3$, $N=256$.}
	\label{upa_NMSE_SNR_distance}
\end{figure}
Fig. \ref{upa_NMSE_SNR_distance} represents the NMSE distance estimation performance of all algorithms as a function of SNR. It is clear that the proposed SG-VB method consistently achieves superior performance compared to other approaches across the entire SNR range. Particularly, the proposed method achieves 8 dB and 8 dB NMSE gain at high SNR in comparison with P-SOMP, SBL, and 3D-MUSIC, respectively. Unlike the ULA case, 3D-MUSIC exhibits poor distance estimation accuracy because it relies on estimated DoAs for distance estimation, making it vulnerable to error propagation.

\section{Conclusions}
In this paper, we developed an efficient SG-VB algorithm for near-field channel estimation in XL-MIMO systems. By introducing new representations, the DoAs and distance were separated into distinct random variables within the estimation problem. Based on these representations, we derived closed-form variational distributions for each random variable under the VB framework, making the approach applicable to both ULA and UPA. To effectively address the challenges posed by near-field spherical-wavefront characteristics, we further proposed a gridless DoA estimation method combined with a refined grid-search strategy for distance estimation. Extensive simulation results demonstrated that the proposed SG-VB algorithm outperforms MUSIC, P-OMP, P-SIGW, SBL, and LS in near-field channel estimation. SG-VB achieved the lowest NMSE in channel reconstruction, distance estimation, and DoA estimation, particularly at high SNRs. Our findings suggest that SG-VB can play a key role in advancing channel estimation for next-generation XL-MIMO systems, contributing to reliable and high-performance wireless communication.

\appendices
\section{Some Results on the von Mises Distribution}\label{appendix:vm}
Given $\theta$ follows a von Mises distribution with mean $\mu$ and concentration $\kappa$, we have
\begin{equation}
    p(\theta) =\frac{\e^{\kappa\cos(\theta-\mu)}}{2\pi I_0(\kappa)}\propto \exp\big\{\Re\{\eta^*\e^{\ji \theta}\}\big\},
\end{equation}
where $\eta = \kappa\e^{\ji \mu}$ and $I_0(\kappa)$ is the Bessel function of the first kind, defined as
$$I_0(\kappa) = \frac{1}{2\pi} \int_{-\pi}^\pi\e^{\kappa \cos x}\mathrm{d}x.$$
We also have the result
$$I_n(\kappa) = \frac{1}{\pi}\int_{0}^\pi\e^{\kappa \cos x}\cos nx\mathrm{d}x.$$

We have circular moments
\begin{align}
    \mathbb{E}\big[\e^{\ji n\theta}\big] &= \frac{I_n(\kappa)}{I_0(\kappa)} \e^{\ji n \mu} = d_n(\kappa)\e^{\ji n\mu} \\
    \mr{var}\big[\e^{\ji n\theta}\big] &= \mathbb{E}\big[|\e^{\ji n\theta}|^2\big] - \big|\mathbb{E}\big[\e^{\ji n\theta}\big]\big|^2 =   1 - \left(\frac{I_n(\kappa)}{I_0(\kappa)}\right)^2.
\end{align}

\def\baselinestretch{.97}
\bibliographystyle{IEEEtran}
\bibliography{Refs_NearField_}

\end{document}